\documentclass[11pt]{article}

\usepackage[usenames,dvipsnames,table]{xcolor}

\usepackage{jheppub} 

\usepackage{graphicx}
\usepackage{amsmath, amssymb}
\usepackage{youngtab}
\usepackage{manfnt}
\usepackage{float}
\usepackage{placeins}
\usepackage{ytableau}
\ytableausetup{boxsize=.6em,centertableaux}
\usepackage{multirow}
\usepackage{arydshln}
\usepackage{dsfont}
\usepackage{slashed}

\def\be{\begin{eqnarray}}
\def\ee{\end{eqnarray}}
\newcommand{\nn}{\nonumber}
\newcommand\para{\paragraph{}}

\newcommand{\eqn}[1]{(\ref{#1})}

\def\Dslash{\,\,{\raise.15ex\hbox{/}\mkern-12mu D}}
\def\Dbarslash{\,\,{\raise.15ex\hbox{/}\mkern-12mu {\bar D}}}
\def\delslash{\,\,{\raise.15ex\hbox{/}\mkern-9mu \partial}}
\def\delbarslash{\,\,{\raise.15ex\hbox{/}\mkern-9mu {\bar\partial}}}
\def\pslash{\,\,{\raise.15ex\hbox{/}\mkern-9mu p}}
\def\calDslash{\,\,{\raise.15ex\hbox{/}\mkern-12mu {\cal D}}}

\newcommand{\ra}{\rightarrow}

\newcommand{\Z}{\mathds{Z}}

\newcommand{\spina}{{\rm Spin}(8)}
\newcommand{\antic}{\raisebox{-1.1ex}{\epsfxsize=0.12in\epsfbox{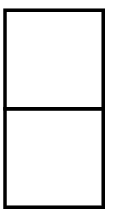}}}

\newcommand{\sym}{\ydiagram{2}}
\newcommand{\anti}{\ydiagram{1,1}}

\newcommand{\la}{\lambda}
\newcommand{\onov}[1]{\frac{1}{#1}}

\def\eqn{\eqref}

\newcommand{\eq}[2][ ]{\begin{equation}\label{#1}{\begin{split}#2\end{split}}\end{equation}}

\title{Chiral Gauge Dynamics: Candidates for Non-Supersymmetric Dualities}

\author{Avner Karasik, Kaan \"{O}nder and David Tong}

\affiliation{Department of Applied Mathematics and Theoretical Physics \\ University Cambridge, CB3 0WA, UK}

\emailAdd{avnerkar@gmail.com, ko354@cam.ac.uk, d.tong@damtp.cam.ac.uk}

\abstract{We study the dynamics of chiral $SU(N)$ gauge theories. These contain Weyl fermions in the symmetric or anti-symmetric representation of the gauge group, together with further fermions in the fundamental and anti-fundamental. We revisit an old proposal of Bars and Yankielowicz who match the 't Hooft anomalies of this theory to free fermions. We show that there are novel and, in some cases, quite powerful constraints on the  dynamics in  the large $N$ limit.

 In addition, we study these $SU(N)$ theories with an extra Weyl fermion transforming in the adjoint representation.  Here we show that all 21  't Hooft anomalies for global symmetries are matched with those of a Spin(8) gauge theory. This suggests a non-supersymmetric extension of the duality of Pouliot and Strassler. Finally, we also discuss some non-supersymmetric dualities with vector-like matter content for $SO(N)$ and $Sp(N)$ gauge theories and the constraints imposed by Weingarten inequalities.}

\begin{document}

\maketitle
\flushbottom

\section{Introduction}

Supersymmetry affords exquisite control over the dynamics of 4d gauge theories. However, for the most part this control has not extended to non-supersymmetric theories.

Indeed, without supersymmetry our understanding of even the most  qualitative questions remains lacking. For example, it has long been known that supersymmetric theories exhibit a broad array of interesting dynamics, including confinement without chiral symmetry breaking, dual theories flowing to the same critical point, and the emergence of  gapless,  ``magnetic" gauge bosons in the infra-red \cite{seib1,seib2}. But it is not yet established whether any of these phenomena can arise in the absence of supersymmetry. (Proposals do exist in the literature, some of which we will describe in more detail below.)

In this paper, we explore some scenarios for the dynamics of certain chiral gauge theories. These chiral theories have gauge group $SU(N)$ and  contain a Weyl fermion in either the symmetric or anti-symmetric representation, together with a collection of further fermions in the fundamental, anti-fundamental and, in some cases,  adjoint representations. These theories do not enjoy supersymmetry and we do not have enough control over their dynamics to be confident of their low-energy physics. We do, however, have a number of tools that constrain the possible dynamics, including 't Hooft anomaly matching, large $N$ techniques and the a-theorem. Using these, we give consistent and, we believe, plausible proposals for the dynamics of these theories that exhibit each of the phenomena described above.

Some of these proposals are not new. In fact, the idea that chiral gauge theories may confine without breaking their global symmetry group predates our understanding of supersymmetric theories \cite{raby}. In particular, Bars and Yankielowicz made a striking proposal that a certain chiral gauge theory confines, with the 't Hooft anomalies saturated by a natural collection of massless composite fermions \cite{by}. Subsequently, some doubt was cast on this suggestion by a study of the theory at large $N$, where it was argued that one of these composite fermions could not remain massless in the limit $N\ra \infty$ \cite{zepp}. We devote much of Section \ref{bysec} to this proposal of Bars and Yankielowicz and revisit what happens in the large $N$ limit. By adapting large $N$ arguments of Coleman and Witten \cite{cw}, we show that in the simplest cases the theory must confine without breaking its global symmetry.  Moreover, we will show that the interpretation of the results of \cite{zepp}, which rule out this behaviour in general, is incorrect and the idea that these theories flow to a collection of free, massless composite fermions remains compelling for certain values of the number of colours $N$ and the number of flavours. We use the a-theorem to give estimates for when this may occur.

In Section \ref{chiralsec}, we turn to  a slightly different chiral theory, one that includes an additional Weyl fermion in the adjoint representation. This $SU(N)$ gauge theory has global symmetry group
\be G = SU(q)\times SU(p) \times U(1)^3\nn\ee
There are 21 independent 't Hooft anomalies associated to $G$. This time there is no collection of free fermions that match all 't Hooft anomalies. There is, however,  a  Spin(8) gauge theory that does the job, with the full collection of 't Hooft anomalies matched to matter  sitting in the ${\bf 8}_v$, ${\bf 8}_s$ and ${\bf 8}_c$ representations, together with some gauge singlets. We describe a plausible scenario in which these two theories are dynamically related, with the Spin(8) gauge bosons emerging in the infra-red from the strong coupling of the $SU(N)$ gauge theory. (The opposite RG flow is also possible for certain values of $N$.)  If this scenario is realised dynamically, it would provide  an extension of the supersymmetric duality discovered by Pouliot and Strassler \cite{ps1} but, in the present case, the proposed duality does not admit a supersymmetric completion. 

In Section \ref{4sec}, we move from chiral theories to vector-like theories. Here we have further tools at our disposal. As first stressed in \cite{ofer}, Vafa-Witten theorems and the related Weingarten theorem can be used to rule out certain putative dualities in any theory that has  a positive semi-definite fermion measure. We review these results and extend them. In particular, we look at some detail at the non-supersymmetric dualities proposed in \cite{adi1,adi2}.





\section{The Bars-Yankielowicz Proposal}\label{bysec}

In this section, we study a pair of chiral gauge theories. Each is built around an $SU(N)$ gauge group, with a Weyl fermion in either the symmetric or anti-symmetric representation, together with a collection of further Weyl fermions in the fundamental and anti-fundamentals. 

Despite a great deal of work exploring the properties of these theories, their dynamics is not fully understood and a number of proposals have been put forward. Prominent among them is a particularly simple suggestion of Bars and Yankielowicz \cite{by}, following earlier work \cite{raby}, that the theory flows to free fermions without breaking its global symmetry group. The 't Hooft anomalies of these free fermions match in a rather striking fashion. Further work discussing various aspects of this theory can be found in \cite{appel,konishi,mury1,mury2,us,avner}. 

Attractive as the Bars-Yankielowicz proposal is, there has long been a roadblock to its realisation. In \cite{zepp}, large $N$ techniques were employed in the  study of chiral gauge theories. It was argued that it is not possible to realise the Bars-Yankielowicz free fermion states at large $N$. Here we give a critical review of these arguments and show that this conclusion is not justified: the Bars-Yankielowicz free fermion proposal is perfectly consistent at large $N$. Moreover, we adapt the large $N$ arguments to show that certain composite fermion states must be massless in this limit.

 In Section \ref{byrevsec}, we review the $SU(N)$ chiral gauge theories and the Bars-Yankielowicz proposal and make some obvious remarks about its possible role in the dynamics of these theories. In Section \ref{largensec}, we analyse these theories through the lens of large $N$ and exhibit a number of constrains in this limit. Finally, in Section \ref{byspecsec}, we describe a scenario in which these chiral theories may exhibit a duality, with the symmetric and anti-symmetric both flowing to the same CFT (together with an additional decoupled sector).

\subsection{$SU(N)$ Chiral Gauge Theories}\label{byrevsec}

Throughout this section, we will consider a pair of chiral $SU(N)$ gauge theories: one with a Weyl fermion in the symmetric $\sym$ representation, and the other with a Weyl fermion in the anti-symmetric $\anti$ representation. (All Weyl fermions here, and below, can be taken to be left-handed.) For now, we treat each in turn. Later, we will see that their low-energy dynamics may be related. 

\subsubsection*{Chiral Gauge Theories with a Symmetric}

We start with the $SU(N)$ gauge theory with a single Weyl fermion in the $\sym$. In addition, we have $p$ Weyl fermions in the fundamental $\Box$ representation and a further $q$ Weyl fermions in the anti-fundamental $\overline{\Box}$. To ensure cancellation of the gauge anomaly, we must take
\be q = N+p+4\label{q}\ee
This theory has non-anomalous, global symmetry group
\be G = SU(q) \times SU(p) \times U(1)_1 \times U(1)_2\label{byg}\ee
where we have omitted discrete quotients. (These discrete quotients won't be important for much of our story, although they play a small role in Section \ref{byspecsec} so we relegate their discussion to  Appendix \ref{quotapp}. For the theory with $p=0$, the discrete quotients were previously discussed in \cite{konishi,us}.)

The full collection of fermions and their transformation under the global symmetry group $G$ is
\be
\begin{array}{|c||c|c|c|c|c|c|}
\hline & SU(N) & SU(q) & SU(p) & U(1)_{1} & U(1)_{2}  \\ \hline 
\chi & \sym & \mathbf{1} & \mathbf{1} & -q & -p \\
\psi & \overline{\Box} & \Box & \mathbf{1} & N+2 & 0  \\
\tilde{\psi} & \Box  & \mathbf{1} & \Box & 0 & N+2 \\
\hline
\end{array}\label{symt}\ee
We'll refer to this as the {\it symmetric theory}. 
It is a simple matter to compute the 't Hooft anomalies of $G$. 
The  two non-Abelian 't Hooft anomalies are:
\be {\cal A}[SU(q)^3] = {\cal A}[SU(p)^3]=N\label{byanom1}\ee
In addition, there are ten further anomalies of the form $X^2\cdot U(1)_a $ with $a=1,2$ and $X$ either gravity or one of the factors of $G$. These anomalies are summarised in the following table:
%
%
%
\be
\begin{array}{|c||c|c|}
\hline  {\cal A}& U(1)_1 & U(1)_2 \\
\hline\hline
SU(q)^2 & N(N+2) & 0 \\ \hline
SU(p)^2 & 0 & N(N+2) \\ \hline
U(1)^2_1& Nq[(N+2)^3-\frac{1}{2}(N+1)q^2]  & -\frac{1}{2}N(N+1)pq^2 \\ \hline
U(1)^2_2  &  -\frac{1}{2}N(N+1)qp^2 & Np[(N+2)^3-\frac{1}{2}(N+1)p^2]\\ \hline
{\rm grav}^2  & \frac{1}{2}qN(N+3)  & \frac{1}{2} pN(N+3) \\ \hline\end{array}
\label{byanom2}\ee
where $q$, $p$ and $N$ are not all independent: they are related by $q=N+p+4$.

Long ago, Bars and Yankielowicz proposed that, at least for certain values of $N$ and $p$, this theory may confine, leaving behind a collection of three different massless, composite fermions, each of which is an $SU(N)$ singlet,
\be \rho_A=\psi\chi\psi\ \ \ ,\ \ \ \rho_S =\tilde{\psi}^\dagger\chi\tilde{\psi}^\dagger \ \ \ ,\ \ \ 
 \rho_B = \psi^\dagger\chi^\dagger\tilde{\psi}\label{symfreefermion}\ee
The indices are contracted such that the fermions have the following transformation properties under $G$, 
\be
\begin{array}{|c||c|c|c|c|}
\hline  & SU(q) & SU(p) & U(1)_{1} & U(1)_{2} \\
\hline \rule{0pt}{2.5ex}  
\rho_A & \anti & \mathbf{1} & N-p & -p \\
\rho_S  & \mathbf{1} & \overline{\sym} & -q& -(N+q) \\
 \rho_B & \overline{\Box} & \Box & p+2 & q-2 \\
 \hline
\end{array}
\label{byfreefermion}\ee
Bars and Yankielowicz pointed out that the 't Hooft anomalies of these free fermions reproduce \eqn{byanom1} and \eqn{byanom2}.

\subsubsection*{Chiral Gauge Theories with an Anti-symmetric}

There is a  similar story for a companion theory, consisting of an $SU(\tilde{N})$ gauge group with a Weyl fermion in the anti-symmetric representation. In fact, it is useful to place this fermion in the $\overline{\anti}$ representation. In addition, we have $q$ Weyl fermions in the $\overline{\Box}$ representation and $p$ Weyl fermions in the ${\Box}$.  This time, the gauge anomaly vanishes only if we take
\be p = \tilde{N} + q-4\label{p}\ee
The global symmetry group $G$ is again given by \eqn{byg} and the matter content is summarised by
\begin{equation}
\begin{array}{|c||c|c|c|c|c|c|}
\hline 
\rule{0pt}{2.5ex} & SU(\tilde{N}) & SU(q) & SU(p) & U(1)_{1} & U(1)_{2}  \\
\hline 
\rule{0pt}{3.1ex} 
\chi &  \overline{\antic}& \mathbf{1} & \mathbf{1} & -q & -p  \\
\psi & {\tiny{{{\yng(1)}}}} & {\bf 1} & {\tiny{\overline{\yng(1)}}}    & 0 & \tilde{N}-2   \\
\tilde{\psi} & {\tiny{\overline{\yng(1)}}} &  {\tiny{\overline{\yng(1)}}}  & \mathbf{1}   & \tilde{N}-2 & 0  \\
\hline
\end{array}
\label{antit}
\end{equation}
We refer to this as the {\it anti-symmetric theory}. Formally, the condition \eqn{p} agrees with \eqn{q} if we take
\be \tilde{N} = -N\nn\ee
Moreover, this relation ensures that   the 't Hooft anomalies of the anti-symmetric theory are again given by \eqn{byanom1} and \eqn{byanom2}. In this sense, the anti-symmetric chiral theory \eqn{antit} can be thought of as the continuation of the symmetric theory \eqn{symt} to negative values of $N$. 

Since the 't Hooft anomalies for the anti-symmetric theory are given by \eqn{byanom1} and \eqn{byanom2}, it is natural to think that they too have a free fermion realisation \eqn{byfreefermion}. The only novelty is the identification of the composite fermions $\rho_S$, $\rho_A$ and $\rho_B$ in terms of the UV degrees of freedom. This time the dictionary is
\be \rho_S=\psi\chi\psi\ \ \ ,\ \ \ \rho_A =\tilde{\psi}^\dagger\chi\tilde{\psi}^\dagger \ \ \ ,\ \ \ 
 \rho_B = \psi^\dagger\chi^\dagger\tilde{\psi}\label{rhos}\nn\ee
where the indices are contracted to give the transformation properties shown in \eqn{byfreefermion}.

A better way to think of this collection of different theories is through the lens of the global symmetry $G$. Any global symmetry group, with its attendant 't Hooft anomalies, can be realised in different field theories. The free fermion avatar \eqn{byfreefermion} holds for any $q$ and $p$. In contrast, the symmetric and anti-symmetric chiral gauge theories hold only for certain ranges of $q$ and $p$. The requirement that $N\geq 2$ means that the symmetric theory provides a realisation of the symmetry group only when
\be q \geq p +6\label{qp6}\nn\ee
(For $N=2$, the gauge theory has a larger symmetry group, but this can be broken to $G$ through suitable interactions with scalar fields or through irrelevant six-fermion operators.) In contrast, the anti-symmetric theory provides a realisation of $G$ only when 
\be q \leq p+2\nn\ee
(This time the cases $\tilde{N}=2,3$ and 4 are all special in the sense that they are non-chiral and may have enhanced symmetries.). Note that the pair of chiral theories do not cover all possible values of $q$ and $p$.

The anomaly matching of Bars and Yankielowicz is certainly striking. The obvious question is: are these massless free fermions realised dynamically? 

The honest answer is: we don't know. At stake is the question of whether the global symmetry group $G$ is dynamically broken or survives in the far infra-red. A firm answer to this is beyond our current understanding of strongly interacting chiral gauge theories. If the global symmetry survives, then anomaly matching means that it must flow to some phase with gapless fermions. This could be an interacting CFT, or it could be the free massless fermions described above. As we now review, the latter is a very  plausible scenario for low $p$ (for the symmetric theory) and low $q$ (for the anti-symmetric theory).

\subsubsection*{Confinement without Chiral Symmetric Breaking}

The question of whether the theories confine without breaking chiral symmetry is a little different for the cases of $q=0$ (in the anti-symmetric theory) and $p=0$ (in the symmetric theory) and the cases with both $p,q\neq 0$. Here we briefly summarise what's known, first in the former case and then in the latter, before turning to more details.

The anti-symmetric theory with $q=0$ was first discussed in \cite{raby} where it was proposed that the theory does indeed confine, without breaking the global symmetry group $G$, leaving behind the single massless, composite fermion $\rho_S$. This is compelling in part because of the sheer simplicity of the proposal. Further evidence comes from the fact that the same physics is observed in the Higgs phase, as well as the confining phase, an observation known, rather grandly, as ``complementarity''. The authors of \cite{raby} also discuss the symmetric theory with $p=0$, again proposing that the theory confines, preserving $G$, now with $\rho_A$ massless. 

 For both of the symmetric and anti-symmetric theories, confinement without chiral symmetry breaking remains the simplest dynamical scenario, but it not the only one. It is also possible that the theory develops a condensate and spontaneously breaks $G$. Various proposals are discussed in \cite{mury1,mury2,us}. No doubt other phases are possible. These different candidates cannot be ruled out using anomalies \cite{us}, so the question of which is favoured is a dynamical one. And we do not have good control over this dynamics. 
 
Some help comes from large $N$. A previous  large $N$ study \cite{zepp} argued that the fermions $\rho_S$ (in the anti-symmetric theory) and $\rho_A$ (in the symmetric theory) are obliged to massless. We review this argument in Section \ref{largensec} and provide another (and, we believe, better) argument that reaches the same conclusion in a more direct fashion.  Hence, it appears that at large $N$ the issue is settled: the theory  does indeed confines, leaving behind massless composite fermions. 

What about the symmetric theory with $p>0$ and the anti-symmetric theory with $q>0$, where the full trinity of Bars-Yankielowicz fermions is needed to match anomalies? At first glance, there seems to be no obstacle to realising these massless fermions dynamically, at least for small $p$ (for the symmetric theory) and small $q$ (for the anti-symmetric theory). We still do not have control over the breaking of $G$, but assuming that it is unbroken in the infra-red, the free massless fermion phase is surely an appropriate candidate. 

There is, however, a fly in the ointment. Or, more precisely, in the literature. This time, the authors of \cite{zepp} argued that, in contrast to the $p=0$ and $q=0$ cases,  the Bars-Yankielowicz fermions cannot be realised in the planar diagram expansion, casting doubt on their dynamical viability. We will review this calculation in Section \ref{largensec} where we claim that the conclusion is flawed: there is no large $N$ obstacle to realising the free fermion Bars-Yankielowicz phase and this remains a plausible candidate for the dynamics for suitably small $p$ and $q$.

\subsubsection*{The Conformal Window}

Of course, when the number of additional fermionic flavours becomes large, there is no mystery about the dynamics. The symmetric theory ceases to be asymptotically free when
\be p \geq p_{AF} = \frac{9}{2}N-3\nn\ee
For $p$ slightly below this value, the theory will flow to a weakly coupled Banks-Zaks fixed point, with $G$ unbroken. As $p$ is decreased still further, this CFT is expected to become more strongly coupled and persist in some conformal window
\be p_\star < p < p_{AF}\nn\ee
We define this window to have $G$ unbroken. The lower end of the conformal window is, like for QCD, unknown. Moreover, in contrast to QCD, we do not have the luxury of lattice simulations to guide us. For $p\leq p_\star$, we presumably return to the story above, with either the $G$ broken phase or the free massless fermions as candidates for the low-energy physics.

Although we do not have a good handle on $p_\star$, some insight comes from the a-theorem \cite{cardy,koma}. This  captures the fact that the number of degrees of freedom must decrease under RG through the statement that the ``$a$ anomaly" obeys $a_{UV} > a_{IR}$. (The authors of  \cite{appel} did a similar analysis, but using the free energy as a proxy for the number of degrees in these theories. However, the free energy is not necessarily decreasing under RG.)

 For free theories, the contribution from various fields is (after absorbing a factor of $90(8\pi)^2$), 
\be a = \mbox{\# scalars} + \frac{11}{2}\mbox{\# Weyl fermions} +  62\,\mbox{\# Gauge Fields}\nn\ee
Suppose that the chiral symmetric theory \eqn{symt} flows to the free massless fermions \eqn{byfreefermion}. Then we have
\be \mbox{Symmetric Theory:}\ \ \ a_{UV} - a_{IR} = \frac{135}{2}N^2 - 11p^2 +\frac{11}{2}N - 44p - 95\nn\ee
This flow is only possible if $a_{UV} > a_{IR}$, which puts an upper bound on $p$. For even reasonably small values of $N$ and $p$, the $N^2$ and $p^2$ terms dominate giving an approximately linear bound
\be p \lesssim 2.48 N\nn\ee
Under the assumption that the theory exits the conformal window to the free fermion phase, this puts an upper bound on the lower end of the conformal window\footnote{For what it's worth, when applied to Seiberg duality of $SU(N_c)$ SQCD with $N_f$ flavours, the a-theorem gives a bound on the lower end of the conformal window of $N_f\lesssim 1.85 N_c$. The true value is, of course, $N_f=1.5 N_c$.}
\be p_\star \lesssim 2.48 N\nn\ee
Again, there is a similar story for  the $SU(\tilde{N})$ anti-symmetric theory. This ceases to be asymptotically free when 
\be q \geq q_{AF} = \frac{9}{2}N +3\nn\ee
Now the a-theorem requires
\be \mbox{Anti-Symmetric Theory:}\ \ \ a_{UV} - a_{IR} = \frac{135}{2}
\tilde{N}^2 - 11q^2 -\frac{11}{2} \tilde{N} +44 q - 95\nn\ee
Again, for large(ish) $N$, the a-theorem tells us that we can only flow to the free-fermion phase when  $q\lesssim 2.48 \tilde{N}$. Again, this could be viewed as an upper bound for the end of the conformal window.

\subsection{Large N}\label{largensec}

The large $N$ expansion is one  of the most important theoretical tools in the analysis of strongly interacting, 4d non-Abelian gauge theories \cite{thooft,witten}. The purpose of this section is to apply large $N$ techniques to  the pair of chiral gauge theories \eqn{symt} and \eqn{antit}. A previous study of these theories at large $N$ was performed by Eichten et al \cite{zepp}. We will explain their results, disagree with some of them, and provide some new ones. We start this section with a review of large $N$ in general before we turn to the chiral theories of interest.

\subsubsection*{A Review of Large $N$ QCD}

To start, consider pure $SU(N)$ gauge theory with coupling $g^2$ and 't Hooft coupling $\lambda=g^2N$. The Yang-Mills action can be written as
\be S_{YM} =\frac{N}{2\la}\int d^4x \ {\rm tr}\,F_{\mu\nu}F^{\mu\nu}\nn\ee
with $F_{\mu\nu}=\partial_\mu A_\nu-\partial_\nu A_\mu-i[A_\mu,A_\nu]$.  The large $N$ limit consists of taking $N\ra \infty$ with $\lambda$ held fixed.

\begin{figure}[h]
	\centering
	\includegraphics[width=0.9 \linewidth]{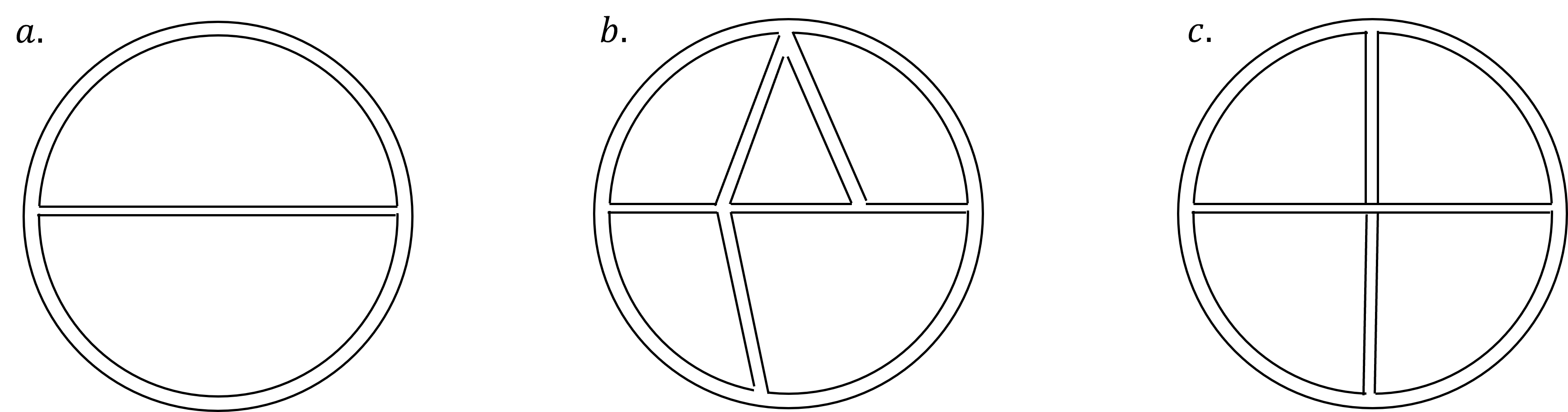}
	\caption{Three diagrams contributing to the vacuum energy of Yang-Mills. Diagrams $a$ and $b$ are both planar and scale like $N^2$. Diagram $a$ has 2 vertices, 3 propagators, and 3 colour loops and so scales as $N^2N^{-3}N^3=N^2$. Diagram $b$ has 6 vertices, 10 propagators, and 6 colour loops resulting again in $N^6N^{-10}N^6=N^2$ scaling. In contrast, diagram $c$ is non-planar: it has 4 vertices, 6 propagators, and 2 colour loops, resulting in $N^4N^{-6}N^2=N^0$ and so is suppressed in the large $N$ limit.}
	\label{vacfig}
\end{figure}

The Feynman rules can be  read from this action. Every gluon propagator is proportional to $1/N$ while every vertex is proportional to $N$ (ignoring the $\la$ dependence).  Feynman diagrams are drawn using the double line representation for the propagators, where each line carries a single colour index. We can then count the additional powers of $N$ that come from summing colour indices: they are equal to the number of loops.  The famous observation of 't Hooft is that the leading contribution to any correlation function comes from planar diagrams. Examples of the contribution to the vacuum energy are shown in Figure \ref{vacfig}.

\begin{figure}[h]
	\centering
	\includegraphics[width=0.9 \linewidth]{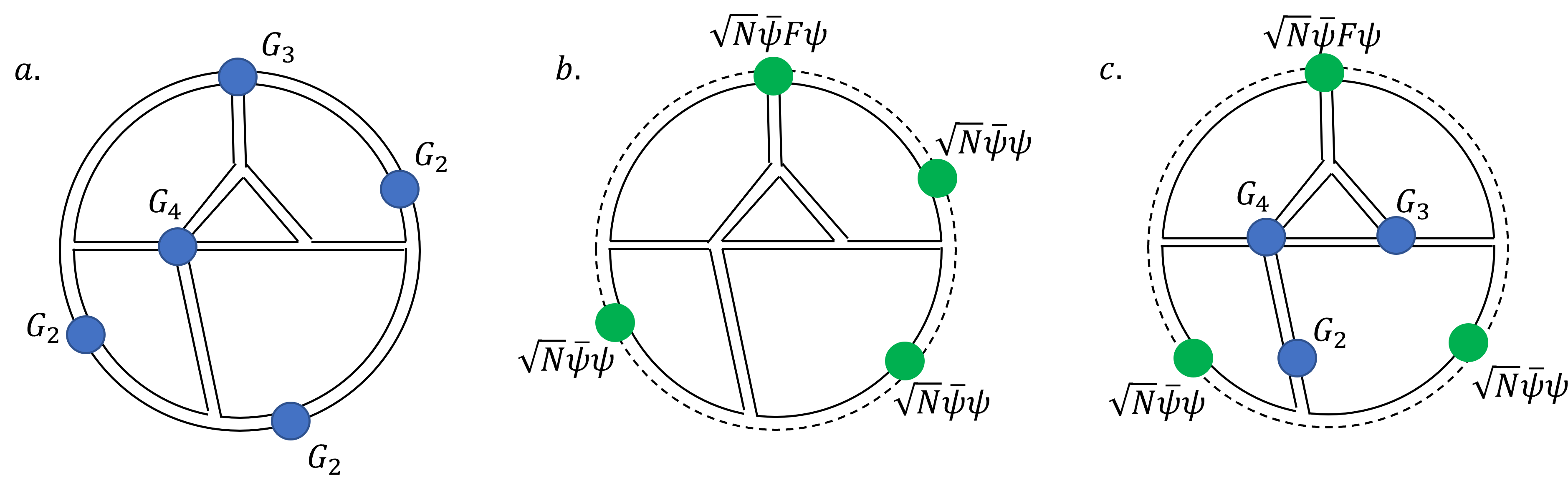}
	\caption{Diagrams with insertions of glueballs (in blue) and mesons (in green). The solid and dashed lines represent colour and flavour indices respectively. Every insertion reduces the number of vertices by 1 or increases the number of propagators by 1. In any case, this gives a factor of $1/{N}$. In addition, a meson insertion comes with a factor of $\sqrt{N}$ due to the normalization.
				This gives the rules described in the text: an $n$-point function of glueballs scales as $N^{2-n}$, as shown in diagram $a$; an $n$-point functions of  mesons scales as  $N^{1-n/2}$, as shown in in diagram $b$; and a mixed correlator of $n_G$ glueballs and $n_M$ mesons scales as $N^{1-n_G-n_M/2}$, as shown in diagram $c$. }
\label{gluecorr}
\end{figure}

Local operators in this theory are glueballs, which we write schematically as 
\be {G}\sim {\rm tr}\,F^k\nn\ee
with $k\geq 2$. The normalisation of these operators, with no explicit factors of $N$,  ensures that the 2-point function scales as
\be  \langle {G(x)G(y)} \rangle \sim {\cal O}(1)\nn\ee
Higher order $n$-point functions then scale as $N^{2-n}$. An example is depicted in the left hand diagram in  Figure \ref{gluecorr}.

Next we add matter to our theory. We start by considering vector-like matter, resulting in a large $N$ version of QCD. This consists of  $N_f$ Weyl fermions, each transforming in the $\Box$ representation of $SU(N)$, and a further $N_f$ Weyl fermions transforming in the $\overline{\Box}$. We denote these fermions as $\psi$ and $\tilde{\psi}$ respectively and pair them together into $N_f$ Dirac fermions $\Psi$. The action is 
\be S_{\rm Dirac} =  N \int d^4x\  \sum_{i=1}^{N_f} i\bar{\Psi}_i\! \Dslash \Psi_i\nn\ee
The overall factor of $N$ means that fermion propagators, like their gluonic counterparts,   scale as $1/N$ while gluon-fermion vertices carry a factor of $N$. The key difference between these fermions and the gluons is that the former carry a single colour index $a=1,\ldots, N$ and a single flavour index $i=1,\ldots N_f$, while the latter carry two colour indices. This has two consequences. The first is that a fermion loop is weighted by a factor of ${N_f}/{N}$ relative to a gluon loop. If $N_f\ll N$ fermion loops will be suppressed. This is referred to as the {\it 't Hooft limit}. If, however,  $N_f\sim N$ then  there is no such suppression: this is referred to as the {\it Veneziano limit} \cite{ven} and will be relevant for the chiral theories that we discuss shortly.

The second consequence is that external states carry fixed flavour index and so these are not summed over. To compute the scaling, we define the meson operators
\be  M \sim \sqrt{N}\tilde{\psi}F^k\psi\label{rootn}\ee
The $\sqrt{N}$ normalization is again chosen so that the two-point function scales as
\be \langle M(x) M(y) \rangle \sim  {\cal O}(1)\label{meson}\nn\ee
Higher order $n$-point correlators of mesons then scale as $N^{1-n/2}$. An example is depicted in the middle diagram in Figure \ref{gluecorr}.

 Using these results, we can determine  the stability of massive states. Glueballs decay as $G\ra GG$ or $G\ra MM$, both of which scale as $1/N$. This gives a  rate of ${\rm diagram}^2 \sim 1/N^2$ for the decay to gluons or, after summing over all possible decay products, $\sim N_f^2/N^2$ for the decay to mesons. If we are working in the limit of fixed $N_f$, this results the well-known statement that the strict $N\ra \infty$ limit enjoys an infinite tower of stable, massive glueball states. If, however, $N_f\sim N$, then glueballs decay with a rate of ${\cal O}(1)$.

 \begin{figure}[h]
	\centering
	\includegraphics[width=0.9 \linewidth]{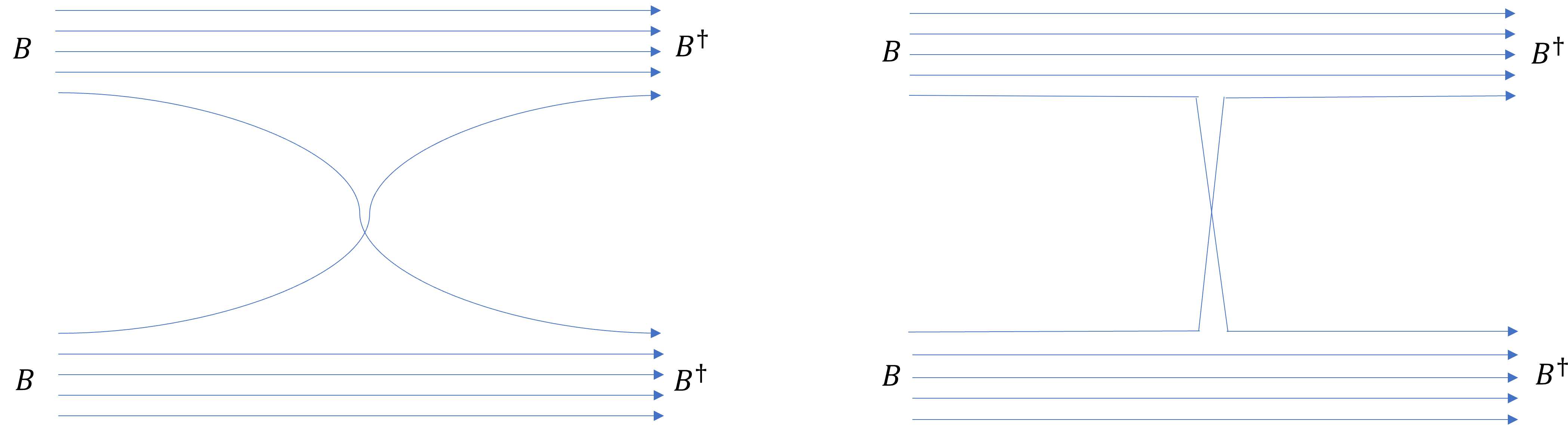}
	\caption{Leading 4-baryon diagrams. On the left diagram the baryons exchange a quark without any interactions. The choice of colour gives a factor of $N$. On the right diagram the baryons exchange a gluon. The choice of colour now gives a factor of $N^2$, the gluon propagator and two vertices give $1/{N}$, resulting again in a total factor of $N$.  }
\label{baryons}
\end{figure}
\noindent

For mesons,  the decay $M\ra MM$ is determined by the diagram $\langle {M_{ij}(x_1)M_{ik}(x_2)M_{kj}(x_3)}\rangle \sim 1/{\sqrt{N}}$. Squaring and summing over all the external states results in a decay rate $\sim {N_f}/{N}$ (because one flavour index is summed over). Again, if $N_f$ is held fixed in the large $N$ limit then there is an infinite tower of stable mesons, while if $N_f\sim N$ then then excited  meson states,  like the glueballs, decay with an ${\cal O}(1)$ rate.

Large $N$ QCD also has a spectrum of baryons that require special attention \cite{witten}. A baryon is an operator whose colours indices are contracted using the epsilon tensor,
\be B^{i_1...i_N}\sim \epsilon^{a_1a_2...a_N}\psi_{a_1}^{i_1}\psi_{a2}^{i_2}\ldots\psi_{a_N}^{i_N}\nn\ee
 A result that we will need later concerns the strength of 4-baryon interaction. As shown by Witten \cite{witten}, the leading contribution to this interaction comes from diagrams in Figure \ref{baryons} and scales as
 \be\langle B(x_1)B^\dagger(x_2) B(x_3)B^\dagger(x_4)\rangle \sim {\cal O}(N)\label{baryon}\ee

Finally, we review the possibilities for symmetry breaking in large $N$ QCD, following the work of Coleman and Witten \cite{cw}. The global symmetry of this theory includes the factor
\be SU(N_f)_L\times SU(N_f)_R\nn\ee
under which the meson operator $M_i^j=\sqrt{N}\tilde{\psi}_i\psi^j$ transforms in the bi-fundamental representation.  This operator provides an order parameter for symmetry breaking. The Coleman-Witten theorem restricts the kind of symmetry breaking that is possible in the large $N$ limit.

For this analysis, it is important to assume that $N_f\ll N$. As described above, the $2n$-point meson interaction is dominated by single trace diagrams and scales as $\langle (M^\dagger M)^n\rangle \sim N^{1-n}$. The  leading terms in the effective potential for the meson then takes the form
\be V(M)=N\sum_n N^{-n}a_n{\rm tr}(M^\dagger M)^n\label{V}\ee
with $a_n\sim {\cal O}(1)$. We can use the  $SU(N_f)_L\times SU(N_f)_R$ transformation to diagonalize the meson operator, so  $M_i^j=\la_i\delta_i^j$. The potential now takes the form,
\be V(\la_i)=N\sum_n N^{-n}a_n \sum_i |\la_i|^{2n} \label{cwhere}\ee
The key observation is that the $\la_i$ are decoupled: each has the same equation of motion, solved by $\la_i=\la$. Therefore, $\langle{M}_i^j\rangle=\la \delta_i^j$ and the only possible symmetry breaking in the large $N$ limit is  $SU(N_f)_L\times SU(N_f)_R\to SU(N_f)_{diag}$. This is the Coleman-Witten theorem \cite{cw}.

The story above holds when $N_f/N\ll 1$. What happens when  $N_f\sim N$? In this case multi-trace operators can become as important as the single trace ones. As an example, we will compare the single-trace and double-trace $(MM^\dagger)^2$ operators,
\be V_2(M) &=&\frac{1}{N} a_2{\rm tr}(MM^\dagger MM^\dagger)+\frac{1}{N^2}b_2{\rm tr}(MM^\dagger)\,{\rm tr}(MM^\dagger)\nn\\ &=&\frac{1}{N}a_2\sum_{i=1}^{N_f}|\la_i|^4+\frac{1}{N^2}b_2\sum_{i,j=1}^{N_f}|\la_i\la_j|^2\label{V2}\ee
The second term is suppressed by $1/N$ relative to the first, but has a double sum over flavour. So if $N_f\sim N$ the two terms are equally important. In this case, the eigenvalues are not decoupled and other symmetry breaking pattern are, in principle, allowed.

\subsubsection*{Large $N$ for Chiral Theories}

After this review, we turn to the large $N$ limit of the chiral theories \eqn{symt} and \eqn{antit}. Our goal is to shed some light on the dynamics of these theories.  A large $N$ analysis of these theories was previously given in \cite{zepp} and we review some of their results below. We focus on the symmetric chiral theory \eqn{symt}. All results carry over immediately to the anti-symmetric theory. 

The fermion $\chi$ transforms in the two-index representation $\sym$ and so, in the large $N$ limit, behaves much like the gluons in the adjoint representation. In particular, loops of $\chi$ in planar diagrams are unsuppressed. 
 The only minor difference is that $\chi$ carries two fundamental indices, in contrast to the gluon which carries one fundamental and one anti-fundamental index. This can be accounted for by adding arrows to propagators and ensuring that they are matched correctly.  

In addition, our theory has $q$ fermions $\psi$ in the $\overline{\Box}$ representation and $p$ fermions $\tilde{\psi}$ in the $\Box$. Anomaly cancellation \eqn{q} means that we necessarily have $q\sim {\cal O}(N)$. Our interest in this section will focus on the regime in which $p\sim {\cal O}(1)$. In this sense, the large $N$ limit of the chiral theory shares some features with both the 't Hooft and Veneziano limits of QCD.

The Bars-Yankielowicz proposal is that the chiral gauge theory confines, resulting in the collection of massless fermions $\rho_S$, $\rho_A$ and $\rho_B$ given in \eqn{symfreefermion}. Each of these composite fermions behaves like a meson in QCD. Including the appropriate meson normalisation \eqn{rootn}, we should replace \eqn{symfreefermion} with
\be \rho_A=\sqrt{N} \psi\chi\psi\ \ \ ,\ \ \ \rho_S =\sqrt{N}\tilde{\psi}^\dagger\chi\tilde{\psi}^\dagger \ \ \ ,\ \ \ 
 \rho_B = \sqrt{N}\psi^\dagger\chi^\dagger\tilde{\psi}\nn\ee
The additional $\sqrt{N}$ factors ensure that the two-point functions are correctly normalised: $\langle \rho(x)\rho(y)\rangle \sim {\cal O}(1)$, with higher point functions suppressed at large $N$.  All other meson-like operators, in which a string of $F$ and $\chi$ fields are sandwiched between some a pair of $\psi$ or $\tilde{\psi}$ fundamentals, are similarly normalised with a $\sqrt{N}$. 

Although higher point functions are suppressed,  excited mesons can still decay with rate of order ${\cal O}(1)$ because the number of $\psi$ flavour is $q\sim {\cal O}(N)$ and these should be summed over in final states. For example, $\psi_i\tilde{\psi}^I$ can decay to $(\rho_A)_{ij}(\rho_B)^{jI}$ where $i,j$ are $SU(q)$ indices and $I$ is an $SU(p)$ index. The probability to decay to a fixed $j$ is $\sim \left(1/{\sqrt{N}}\right)^2=1/{N}$. Summing over $j$ we get an ${\cal O}(1)$ decay rate.

\subsubsection*{The $p=0$ Theory}

We start by looking at the $p=0$ theory, where there are $q=N+4$ $\psi$ fields and no $\tilde{\psi}$ fields. In this case, the global symmetry group is just
\be G = SU(q) \times U(1)\nn\ee
and the conjectured dynamics is that the theory confines, preserving $G$, resulting in a single gapless fermion $\rho_A$. 

The authors of \cite{zepp} studied this theory in the large $N$ limit and provided an, admittedly slightly circuitous, argument that $\rho_A$ must indeed be massless. Their argument proceeds by showing that the global $U(1) \subset G$ cannot be spontaneously broken in the large $N$ limit because the associated Goldstone bosons cannot be seen  in cuts of planar diagrams\footnote{This argument was recently criticised in \cite{yetagain} by working with non-gauge invariant operators in the planar limit. We do not agree with this criticism.}. The $U(1)$ 't Hooft anomalies mean that there must be $\frac{1}{2}q(q-1)$ massless fermions, and it would be perverse if these did not sit in the anti-symmetric representation of $SU(q)$. 

Here we give an alternative, more direct argument that shows the $\rho_A$ fermion must be massless in the large $N$ limit. To see this, first note that, by anomaly matching, if the fermion $\rho_A$ is not massless then the symmetry $G$ must be spontaneously broken. We introduce a gauge invariant  order parameter $\Phi$ for this symmetry breaking. The fermion gets its mass through interactions with the condensate $\Phi$, which can then be determined by computing an appropriate correlation function, such as  $\langle \Phi \rho_A\rho_A\rangle$. To determine how the mass scales with $N$, we need to compute both the scaling of these correlation functions, and the scaling of the expectation value of $\langle \Phi\rangle$. Both are straightforward.

\para
It's useful to think of the correlation function $\langle \Phi \rho_A\rho_A\rangle$ as giving rise to an effective Yukawa interaction. The most general form is
\be{\cal L}_{\rm Yuk} \sim y\Phi_{ijkl} \,\rho_A^{ij}\rho_A^{kl} + {\rm h.c.}\label{yuk}\ee 
with $i,j,k,l=1,\ldots q$. Here the Yukawa coupling $y$ will have some scaling on $N$ determined by $\langle \Phi\rho_A\rho_A\rangle$. The exact form of this scaling depends on the choice of operator $\Phi$ so the next question is: what kind of gauge invariant operators $\Phi$ can we construct?

First note that there is no single trace meson-like operator with 4 $SU(q)$ indices. But we can construct ``multi-trace" (really, ``multi-meson" in this context) operators and we can construct baryonic operators. We start with the former.

An example of a multi-trace operator is simply $\Phi = \rho_A^\dagger \rho_A^\dagger$. There is nothing to prohibit such an operator getting an expectation value of, say, the form
 \be \langle \Phi_{ijkl}\rangle=v[\delta_{ik}\delta_{jl}- \delta_{il}\delta_{jk}] \nn\ee
which has the ability to gap the fermion $\rho_A$. From the general rules described previously, we know that the Yukawa coupling in this case must scale as 
$y\sim 1/N$. This,  in turn, means that if the mass in \eqn{yuk}  is to survive in the large $N$ limit, then the expectation value $v$ would have to scale as $v\sim N$. But this is forbidden. We can see this by looking at the effective potential \eqn{V2} for $\Phi$. This includes terms
\be V_2(\Phi,\Phi^\dagger) \sim \frac{1}{N} \Phi^{\dagger\,ijkl}\Phi_{ijkl} \sim \frac{q^2v^2}{N}\nn\ee
If $v\sim N$, then this contribution to the vacuum energy would scale as $V_2 \sim q^2 N \sim N^3$. But that's illegal! The vacuum energy must scale as $N^2$. We learn that  $v\sim {N}$ is not allowed and, from \eqn{yuk}, the fermionic mass is suppressed by $1/N$. The double-trace operator $\Phi= \rho_A^\dag\rho_A^\dag$ cannot gap the fermions in the large $N$ limit. 

The same conclusion is reached if we consider other multi-meson operators. But what about baryons? For example, we can construct a scalar operator of the form $\Phi_B \sim \chi^{N-4}\psi^{N-8}$, with the hanging colour indices contracted with an $SU(N)$  epsilon symbol. Is it possible to gap the fermions through some expectation $\langle \Phi_B\rangle$? The answer, again, is no. From \eqn{baryon}, the effective potential for $\Phi_B$ scales as
\be V(\Phi_B^\dagger,\Phi_B) \sim |\Phi_B|^2 + N |\Phi_B|^4 + \ldots
\nn\ee
and so any expectation value must scale as $\langle \Phi_B\rangle \sim 1/\sqrt{N}$. The Yukawa term \eqn{yuk} then gives a mass for the fermions that scales as ${\rm mass} \sim y/\sqrt{N}$. This mass vanishes in the large $N$ limit provided that $y$ does not grow with $N$.  This is surely the case: the vertex $\Phi_B\rho_A\rho_A$, with $\Phi_B$ a baryon, vanishes in perturbation theory because the total charge under the anomalous $U(1)$ is non-zero. Such a vertex should emerge non-perturbatively but is certainly  expected to be suppressed in the large $N$ limit.

The upshot is that the fermion $\rho_A$ must be massless in the large $N$ limit.

The same arguments go through for the anti-symmetric theory with $q=0$. In this case, the fermion $\rho_S$ must be massless in the large $N$ limit. 

\subsubsection*{The $p>0$ Theories}

We can apply the same kind of arguments to the Bars-Yankielowicz proposal that all three composite fermions, $\rho_A$, $\rho_S$ and $\rho_B$ remain massless. Importantly, we work in the limit $p/N\ra 0$ and $N\ra \infty$. This time, as we shall see, the results are not so strong. There is  no obstacle to a collection of these fermions becoming gapped, with the commensurate symmetry breaking. However, in contrast to \cite{zepp}, we will argue that nor is there any reason to believe that the fermions are necessarily gapped.

First, we repeat the arguments above and show that they now allow for the possibility  that the fermions are gapped in the large $N$ limit. This is achieved through the scalar operator
\be \Phi_{iI}=\sqrt{N}\psi_i\tilde{\psi}_I\nn\ee
with $i=1,\ldots,q$ and $I=1,\ldots,p$. The interactions with the mesons are captured by the Yukawa interactions
\be {\cal L}_{\rm Yuk} \sim \frac{1}{\sqrt{N}} y_{A} \Phi^\dagger_{iI}(\rho_A)^{ij}(\rho_B)_j^I+\frac{1}{\sqrt{N}}y_{S}\Phi^{ iI}(\rho_S)_{IJ}(\rho_B)_i^J+ {\rm h.c}\label{yukky}\ee
with $y_{S,A}\sim 1$. The overall $1/{\sqrt{N}}$ comes from the scaling of three-meson diagrams. 

The expectation value of $\Phi$ can be determined by looking at the effective potential. In the large $N$ limit, this is given by a sum of single trace terms as in \eqn{V}. (The multi-trace terms that arose in the Veneziano limit in \eqn{V2} are once again suppressed here because  ${\rm rank}(\Phi) \leq p\ll N$.) This tells us that any expectation value scales as $\langle\Phi\rangle \sim\sqrt{N}$. Substituting this back into the potential gives an entirely expected, and entirely legal, contribution to the vacuum energy $V\sim Np$. Meanwhile, substituting this expectation value into \eqn{yukky} gives a collection of the fermions a mass that is unsuppressed in the large $N$ limit.

The fact that single trace operators dominate in the potential for $\Phi$ means that a version of the Coleman-Witten theorem applies. After diagonalisation by $SU(q)\times SU(p)$, we can write the scalar in the form
\be
\Phi=\left(\begin{array}{ccc}
\lambda_{1} & \ldots & 0 \\
& \ddots & \\
0 & \ldots & \lambda_{p} \\
\hdashline 0 & \ldots & 0
\end{array}\right)
\nn\ee
The potential then splits into independent terms for each $\lambda_i$, as in \eqn{cwhere}. This tells us that each $\lambda_i$ takes the same value in the ground state, and the symmetry breaking pattern is
\be{SU(q)\times SU(p)\times U(1)_1\times U(1)_2\to SU(q-p)\times SU(p)_{diag}\times U(1)^2\ ,}\label{wearecw}\ee
where $U(1)^2\subset SU(q)\times U(1)_1\times U(1)_2$. The remaining $SU(q-p)$ means that the fermions $(\rho_A)^{ij}$, $i,j=1,...,q-p$ necessarily remain massless, transforming in the anti-symmetric representation of $SU(q-p)$ and saturating the surviving 't Hooft anomaly.

This argument shows that the symmetry broken phase is consistent in the large $N$ limit, but it doesn't whether it is dynamically preferred. We don't know the answer to that. The authors of \cite{zepp} argue something stronger: that it is not possible to realise the full spectrum of massless fermions $\rho_A$, $\rho_S$ and $\rho_A$ in the large $N$ limit. The offending fermion is $\rho_S$, transforming in the $\sym$ of $SU(p)$. This is special because it has a finite number of components in the $N\ra \infty$ limit, in contrast to both $\rho_A$ and $\rho_B$, each of which transforms under $SU(q)$ and so have an infinite number of components as $N\ra \infty$. The argument of \cite{zepp} is that it is not possible to find $\rho_S$ in the cut of a planar diagram. Translated into correlation functions, this is the same statement that the decay of an excited meson  $M^I_J$ into $\rho_S\rho_S^\dagger$ is suppressed in the large $N$ limit
\be \langle M^I_J\,\rho_S^{JK}\,(\rho^\dagger_S)_{IK}\rangle \sim {\cal O}\left(\frac{1}{\sqrt{N}}\right)\label{notwhatyouthink}\ee
This means that the decay rate scales as $\sim p/N$. The lack of such a decay was interpreted as the statement that $\rho_S$ cannot be massless, since then such a decay would be expected to proceed at a rate of order ${\cal O}(1)$. We see no reason to adopt this interpretation. The situation is entirely analogous to pions in QCD.  A generic massive state in the 't Hooft limit does not decay to pions, but this is not because pions are heavy; it is because the interaction is suppressed. 

The same is true of the result \eqn{notwhatyouthink}. It does not restrict the mass of $\rho_S$. Indeed, the same $1/\sqrt{N}$ scaling is seen for appropriate decays to $\rho_A$ and $\rho_B$ as well, but these are unsuppressed after summing over the possible final states. The upshot is that the large $N$ limit does not rule out a phase of gapless fermions with unbroken chiral symmetry $G$. It does, however, predict that if this occurs then the $\rho_S$ sector is decoupled from heavy states.

The argument in this section was restricted to $p\sim {\cal O}(1)$. For $p\sim {\cal O}(N)$, symmetry breaking patterns other than  \eqn{wearecw} may occur. Alternatively, if $G$ is unbroken then $\rho_S$ no longer decouples from the dynamics in the $N\ra \infty$ limit. 

\subsection{A Speculative Duality}\label{byspecsec}

Above, we described various possible phases of the symmetric and anti-symmetric chiral theories. These theories share the same global symmetry group $G=SU(q)\times SU(p) \times U(1)^2$, but exist for complementary values of $q$ and $p$. For this reason, there's no way that the two theories can flow to the same interacting CFT. The purpose of this section is to present a deformation of these theories that may induce them to flow to the same fixed point. We stress that the results that we present here are plausible, but rely on several assumptions about the dynamics that may not be realised in practice.

\subsubsection*{The Symmetric Theory}

We start by returning to the $SU(N)$ symmetric theory, but with the addition of a singlet fermion in the UV. The matter content is
\be
\begin{array}{|c||c|c|c|c|c|c|}
\hline & SU(N) & SU(q) & SU(p) & U(1)_{1} & U(1)_{2}  \\ \hline 
\chi & \sym & \mathbf{1} & \mathbf{1} & -q & -p \\
\psi & \overline{\Box} & \Box & \mathbf{1} & N+2 & 0  \\
\tilde{\psi} & \Box  & \mathbf{1} & \Box & 0 & N+2 \\
 \nu & {\bf 1} & {\Box} & \overline{\Box} & -(p+2) & 2-q \\
\hline
\end{array}\nn\ee
As before, we must have $q=N+p+4$ to ensure consistency of the theory. We now add the 4-fermion term
\be {\cal L}_{\rm 4-fermi} \sim \psi^\dagger \chi^\dagger \tilde{\psi}\nu + {\rm h.c.}\label{4fermi}\ee
This is, of course, an irrelevant operator in the UV. However, it is plausible that, in certain situations, it is a dangerously irrelevant operator, with  the strong coupling dynamics giving it a large anomalous dimension in the infra-red. Since the theory is weakly coupled for large $p$, this can only take place for suitably small values of $p$
\be p<p_{DI}\nn\ee
where $p_{DI}<p_{AF}$. The exact value of $p_{DI}$ is, of course, not known. 

For $p\geq p_{DI}$, the infra-red physics coincides with that in the previous section, now with the additional fermion $\nu$ as a bystander.  But for $p<p_{DI}$, the 4-fermion term drives us to a new fixed point.

If we are in a phase where the theory flows to the free fermions  then it's clear what happens: after confinement, the 4-fermion term \eqn{4fermi} descends to a mass term ${\cal L}_{\rm 4-fermi} \sim \rho_B\nu + {\rm h.c.}$ in the infra-red where $\rho_B$ is the bi-fundamental composite defined in   \eqn{byfreefermion}. Both of these fermions then get a mass and we're left just with the pair of massless fermions $\rho_A$ and $\rho_S$, with charges
\be
\begin{array}{|c||c|c|c|c|}
\hline  & SU(q) & SU(p) & U(1)_{1} & U(1)_{2} \\
\hline \rule{0pt}{2.5ex}  
\rho_A & \anti & \mathbf{1} & N-p & -p \\
\rho_S  & \mathbf{1} & \overline{\sym} & -q& -(N+q) \\
 \hline
\end{array}
\nn\ee
At this point, it is useful to redefine the $U(1)$ charges. We introduce new charges $Q_1'$ and $Q_2'$, related to the previous charges $Q_1$ and $Q_2$ as
\be Q_1'= \frac{N+q}{N+2}Q_1 - \frac{q}{N+2} Q_2\ \ \ {\rm and}\ \ \ Q_2'= \frac{p}{N+2} Q_1+ \frac{N-p}{N+2} Q_2\nn\ee
It's not immediately obvious that this transformation is legal because of the denominator $1/(N+2)$. But one can check that the charges of all UV fermions remain integer. The advantage of this new basis  that it diagonalises the charges of the IR composites, which read
\be
\begin{array}{|c||c|c|c|c|}
\hline  & SU(q) & SU(p) & U(1)'_{1} & U(1)'_{2} \\
\hline \rule{0pt}{2.5ex}  
\rho_A & \anti & \mathbf{1} & 2N & 0 \\
\rho_S  & \mathbf{1} & \overline{\sym} & 0& -2N \\
 \hline
\end{array}
\nn\ee
This reflects the fact that, without the bi-fundamental fermion $\rho_B$, the group structure factorises in the infra-red, with 
\be G = G_1 \times G_2 = [SU(q)\times U(1)_1'] \times [SU(p)\times U(1)'_2]\label{factor}\ee
Here, factorisation is the statement that there are are no mixed 't Hooft anomalies between $G_1$ and $G_2$. 

In fact, a more careful analysis shows that the factorisation is not quite complete: there are discrete quotients that forbid certain representations. To avoid distraction from the main narrative, we describe this in Appendix \ref{quotapp}.

The factorisation \eqn{factor} is  trivial to see from the free fermion perspective, simply because there are no degrees of freedom with charges under both $G_1$ and $G_2$. But it is much more surprising result when viewed from the perspective of the original $U(1)$ theory which does have degrees of freedom charged under both subgroups. (For example, one can check that the $U(1)_1'$ and $U(1)_2'$ charges of $\chi$, $\psi$ and $\tilde{\psi}$ are all non-vanishing.) The existence of the factorisation is an interesting feature of the chiral gauge theory with the four-fermion term. The question that we would like to ask is: does it hold a further clue for the dynamics?

As we've seen, the effect of the dangerously irrelevant operator \eqn{4fermi} is straightforward in the free fermion phase. What might its effect be in the conformal window, assuming that $p_{DI} > p_\star$, so that the 4-fermion term does indeed provide a relevant nudge in the infra-red. It will make the theory flow to a new fixed point and, once again, there are various options: we may flow to the free fermion theory, or break the global group $G$. But we could flow to a different interacting fixed point. And because the global symmetry group and its anomalies factorise, it is natural to postulate that so too does the infra-red CFT, which takes the form of two decoupled sectors
\be {\rm CFT}\Big[q;\anti\Big] \otimes {\rm CFT}\Big[p;\sym\Big]\nn\ee
Here CFT$[q;\anti]$ saturates the 't Hooft anomalies of $G_1 = SU(q)\times U(1)_1'$, which equal those of the anti-symmetric fermion $\rho_A$. Meanwhile, CFT$[p;\sym]$ saturates the 't Hooft anomalies of $G_2=SU(p)\times U(1)_2'$, which equal those of the symmetric fermion $\rho_S$.

\subsubsection*{The Anti-Symmetric Theory}

We can repeat the same story for the anti-symmetric theory. For our purposes, it will be useful to think of the $SU(\tilde{N})$ theory, with global symmetry group
\be \tilde{G} = SU(\tilde{q}) \times SU(\tilde{p})\times U(1)_1\times U(1)_2\nn\ee
We will leave $\tilde{N}$, $\tilde{q}$ and $\tilde{p}$ arbitrary for now; only later will we see how they relate to $N$, $p$ and $q$ of the symmetric theory. We again add a gauge singlet, so the full collection of fermions is
\begin{equation}
\begin{array}{|c||c|c|c|c|c|c|}
\hline 
\rule{0pt}{2.5ex} & SU(\tilde{N}) & SU(\tilde{q}) & SU(\tilde{p}) & U(1)_{1} & U(1)_{2}  \\
\hline 
\rule{0pt}{3.1ex} 
\chi &  \overline{\antic}& \mathbf{1} & \mathbf{1} & -\tilde{q} & -\tilde{p}  \\
\psi & {\tiny{{{\yng(1)}}}} & {\bf 1} & {\tiny{\overline{\yng(1)}}}    & 0 & \tilde{N}-2   \\
\tilde{\psi} & {\tiny{\overline{\yng(1)}}} &  {\tiny{\overline{\yng(1)}}}  & \mathbf{1}   & \tilde{N}-2 & 0  \\
 \nu & {\bf 1} & {\Box} & \overline{\Box} & -(\tilde{p}+2) & 2-\tilde{q} \\
\hline
\end{array}
\nn
\end{equation}
The gauge anomaly cancels only if $\tilde{p} = \tilde{N} + \tilde{q} -4$. The singlet $\nu$ couples to the other fermions through the same (possibly dangerously) irrelevant four fermion term \eqn{4fermi}.

The story is entirely analogous to the symmetric theory. For large $\tilde{q}$, the theory is weakly coupled and the four-fermion term is irrelevant. For small $\tilde{q}$, there is the possibility that it becomes dangerously irrelevant. This happens for some
\be \tilde{q}<\tilde{q}_{DI}\nn\ee
If the free fermion phase is realised, it is again clear that the role  of the 4-fermion interaction \eqn{4fermi} is to gap the bi-fundamental composite $\rho_B$. Here, our interest is in the region of the  conformal window in which the 4-fermion operator is relevant. Under the assumption that it flows again to two decoupled CFTs, we have
\be \widetilde{\rm CFT}\Big[\tilde{q};\anti\Big] \otimes \widetilde{\rm CFT}\Big[\tilde{p};\sym\Big]\nn\ee
At this point, an obvious duality suggests itself. Whenever one of the CFTs arising from the symmetric theory and the anti-symmetric theory share the same symmetry group and anomalies, it is natural to conjecture that they are the same CFT. For the $U(1)$ charges to match, we require
\be N = \tilde{N}\nn\ee
We then conjecture that when $q=\tilde{q}$ or when $p=\tilde{p}$, one of the factors coincide
\be {\rm CFT}\Big[ q;\anti\Big] = \widetilde{\rm CFT}\Big[ q;\anti\Big]\ \ \ {\rm and}\ \ \   {\rm CFT}\Big[ p;\sym\Big] = \widetilde{\rm CFT}\Big[ p;\sym\Big] \nn\ee
Note that it's not possible for both infra-red factors to coincide in the two theories because, for example, $q=\tilde{q} \Rightarrow p \neq \tilde{p}$. This is trivial to see because $q>p$ while $\tilde{p}> \tilde{q}$. Instead, when either $q=\tilde{q}$ or $p=\tilde{p}$, and both theories flow to a pair of CFTs, one of this pair share the same symmetries and anomalies and so are plausibly the same.

\section{Chiral Gauge Theories and Spin(8) Theories}\label{chiralsec}

In this section, we study a slight variant of the chiral $SU(N)$ theories described in the previous section. They differ only by the addition of an extra adjoint fermion. We will show that they share the same global symmetry and 't Hooft anomalies as a non-chiral theory with Spin(8) gauge group. (As we will see, strictly the gauge group must be $\spina\times {\Z}_2$.) 
This anomaly matching is an extension of the supersymmetric duality of Pouliot and Strassler \cite{ps1} but, as we will see, with  matter content that does not admit a supersymmetric completion. Nonetheless, we will explore the possibility that the $SU(N)$ theory flows in the infra-red to the Spin(8) gauge theory.

\subsection{Anomaly Matching}

The theories in this section have global symmetry group
\be G = SU(q)\times SU(p) \times U(1)_1 \times U(1)_2 \times U(1)_R\nn\ee
for some $q$ and $p$.  We start by ignoring discrete quotients of $G$, but they will be important later and will be described in some detail in Section \ref{discretesec}. The name of the $U(1)_R$ group is taken from the supersymmetric context, where it coincides with the R-symmetry of the theory. 

\subsubsection*{Chiral Gauge Theories}

The first theory that realises this global symmetry has $SU(N)$ gauge group and a collection of left-handed Weyl fermions, notably with one transforming in the symmetric $\sym$ representation of $SU(N)$. The full collection transforms as
\be
\begin{array}{|c||c|c|c|c|c|c|}
\hline & SU(N) & SU(q) & SU(p) & U(1)_1 & U(1)_2 & U(1)_R \\
\hline \lambda & \operatorname{adj} & \mathbf{1} & \mathbf{1} & 0 & 0 & N+2 \\
\chi & {\tiny{{\yng(2)}}} & \mathbf{1} & \mathbf{1} &-q & -p & -2 N \\
\psi & {\tiny{\overline{{\yng(1)}}}} & {\tiny{{\yng(1)}}} & \mathbf{1} & N+2 & 0 & 0 \\
\tilde{\psi} & {\tiny{{\yng(1)}}} & \mathbf{1} & {\tiny{{\yng(1)}}} & 0 & N+2 & 0 \\
\hline
\end{array}
\label{syml}\ee
Clearly this is an extension of the symmetric theory \eqn{symt} that we met previously. It differs only by the addition of the adjoint fermion $\lambda$. In the supersymmetric context, $\lambda$ plays the role of the gaugino. There is nothing to prohibit extending this theory to have ${\cal N}=1$ supersymmetry and we briefly describe some of the known results about supersymmetric dynamics below. Nonetheless, we will consider the non-supersymmetric version, motivated by the fact that the Spin(8) theory that shares the anomalies does not have a supersymmetric completion.

As in Section \ref{bysec}, the theory is consistent only if
\be q = N+p+4\label{qpN}\ee
The symmetric chiral theory only provides a realisation of the global symmetry group $G$ for $q \geq p+6$. For $q$ and $p$ outside this range, there is a second, different chiral theory that has the same $G$ and the same anomalies. This theory has gauge group $SU(\tilde{N})$, now with a Weyl fermion in the anti-symmetric $\anti$ rather than symmetric representation. The full collection of left-handed fermions transforms as
\be
\begin{array}{|c||c|c|c|c|c|c|}
\hline 
\rule{0pt}{2.5ex} & SU(\tilde{N}) & SU(q) & SU(p) & U(1)_{1} & U(1)_{2} & U(1)_R \\
\hline
\rule{0pt}{2.5ex} \lambda & \operatorname{adj} & \mathbf{1} & \mathbf{1} & 0 & 0 & 2- \tilde{N} \\
\chi & \overline{\antic} & \mathbf{1} & \mathbf{1} & -q & -p & 2\tilde{N} \\
\psi & {\tiny{{{\yng(1)}}}} &\mathbf{1} & {\tiny{\overline{\yng(1)}}} &  0 & \tilde{N}-2 & 0 \\
\tilde{\psi} & {\tiny{\overline{\yng(1)}}} &  {\tiny{\overline{\yng(1)}}} & {\bf 1}  &  \tilde{N}-2 & 0 & 0 \\
\hline
\end{array}
\label{antil}\ee
Again, this is an obvious extension of the anti-symmetric theory of Section \ref{bysec}, with the addition of the adjoint fermion $\lambda$. 
The theory is consistent if we take
\be p = \tilde{N} + q-4\nn\ee
This coincides with \eqn{qpN} if we formally identify $\tilde{N} = -N$.  This anti-symmetric theory carries a representation of $G$ in the range $q \leq p+2$.

As in Section \ref{bysec}, the identificatation $N=-\tilde{N}$ allows us to write the 't Hooft anomalies of the symmetric and anti-symmetric theories in a unified form. The  two non-Abelian 't Hooft anomalies are again given by \eqn{byanom1}.  In addition, there are 18 further anomalies of the form ${\cal A}[X^2\cdot U(1)_a]$ with $a=1,2,R$ and $X$ either gravity or one of the  global symmetries. These anomalies are summarised in the following table:
\be
\begin{array}{|c||c|c|c|}
\hline  {\cal A}& U(1)_1 & U(1)_2  & U(1)_R\\
\hline\hline
SU(q)^2 & N(N+2) & 0 & 0 \\ \hline
SU(p)^2 & 0 & N(N+2)  & 0 \\ \hline
U(1)^2_1& Nq[(N+2)^3-\frac{1}{2}(N+1)q^2]  & -\frac{1}{2}N(N+1)pq^2 &  -N^2(N+1)q^2  \\ \hline
U(1)^2_2  &  -\frac{1}{2}N(N+1)qp^2 & Np[(N+2)^3-\frac{1}{2}(N+1)p^2] & -N^2(N+1)p^2 \\ \hline
U(1)_R^2  &   -2N^3(N+1)q & -2N^3(N+1)p & \mbox{See Below} \\ \hline
{\rm grav}^2  & \frac{1}{2}qN(N+3)  & \frac{1}{2} pN(N+3)  & (N+1)(N-2)  \\ \hline\end{array}
\ \ \ \ \label{anom2}\ee
The one, missing entry is given by
\be 
{\cal A}[U(1)_R^3] = -(N+1)(N-2)(3N^3+N^2-4N-4) \label{u1anom}\ee
Lastly there is the mixed anomaly involving all three $U(1)$'s,
\be {\cal A}[U(1)_1\cdot U(1)_2\cdot U(1)_R] = -N^2(N+1)pq\label{anom3}\ee

\subsubsection*{A Spin(8) Gauge Theory}

There is a very different gauge theory which realises the same symmetry group $G$ and 't Hooft anomalies. This is, roughly speaking, a $\spina$ gauge theory with the following matter content. (We will explain what  the subtlety hidden in ``roughly speaking" in Section \ref{discretesec}.)
\begin{equation*}
\begin{array}{|c||c|c|c|c|c|c|}
\hline & \spina & SU(q) & S U(p) & U(1)_{1} & U(1)_{2} & U(1)_R \\
\hline \rule{0pt}{2.5ex} 
\tilde{\lambda} & \operatorname{adj}=\mathbf{2 8} & \mathbf{1} & \mathbf{1} & 0 & 0 & N+2 \\
v & \mathbf{8}_{v} & \overline{{\tiny{{\yng(1)}}}} & \mathbf{1} & -\frac{1}{2}(N-p) & \frac{1}{2}p & -\frac{1}{2}(N+6) \\
s & \mathbf{8}_{s} & \mathbf{1} & {\overline{\tiny{{\yng(1)}}}} & -\frac{1}{2}q & -\frac{1}{2}(N+q) &\frac{1}{2}(N+6) \\
c & \mathbf{8}_{c} & \mathbf{1} & \mathbf{1} & \frac{1}{2}N q & \frac{1}{2}Np & \frac{1}{2}N(N-2) \\
\sigma_S & \mathbf{1} & {\tiny{{\yng(2)}}} & \mathbf{1} & N-p & - p & 4 \\
\sigma_A & \mathbf{1} & \mathbf{1} &\overline{\antic} &  -q & -(N+q) & 4 \\
\rho_B & \mathbf{1} & \overline{{\tiny{{\yng(1)}}}} & {\tiny{{\yng(1)}}} & p+2 & q-2 & -4 \\
u & \mathbf{1} & \mathbf{1} & \mathbf{1} & -qN & - pN  & -(N^2-N+2) \\
\hline
\end{array}
\end{equation*}
One can check that all 21  't Hooft anomalies of this theory are again given by \eqn{byanom1}, \eqn{anom2}, \eqn{u1anom} and \eqn{anom3}. 

The $\spina$ theory includes a number of gauge singlets. They must be endowed with their global quantum numbers under $G$ through appropriate interaction terms. These could come about by introducing scalars to the theory, with appropriate Yukawa couplings, or by considering the irrelevant 4-fermion terms
\be
\mathcal{L}_4\sim(c\tilde{\lambda})(cu)+s^\dag s^\dag \tilde{\lambda} \sigma_A + v\tilde{\lambda}v\sigma_S +  \rho_B\sigma_S \rho_B\sigma_A
\nn\ee
with the indices contracted appropriately to impose the desired (anti)-symmetric structure on $\sigma_A$ and $\sigma_S$.

Note that the gauge singlet  $\rho_B$ carries the same quantum numbers as its namesake in the Bars-Yankielowicz free fermion theory. But the quantum numbers of $\sigma_A$ and $\sigma_S$ above are not the same as the Bars-Yankielowicz $\rho_A$ and $\rho_S$: the anti-symmetric under $SU(q)$ is exchanged for a symmetric, and vice-versa for $SU(p)$. The $SU(q)^3$ anomaly for the anti-symmetric representation is $N-4$; for the  symmetric representation it is $N+4$. Roughly speaking,  the Spin(8) theory is an extension of the Bars-Yankielowicz free fermions whose anomaly matching works because $N-4+8=N+4$. However, the matching of all 21 anomalies appears quite non-trivial\footnote{As a note of caution: there are examples in supersymmetric theories where anomaly matching is known to be misleading \cite{mislead}. There are also non-supersymmetric theories whose anomalies match whose dynamical meaning remains unclear  \cite{terning}.}. The main purpose of this section is to explore whether this matching can be realised dynamically, with one theory flowing to the other.

In the case of $p=0$, there is a straightforward answer. On the chiral side, the $p=0$ symmetry can be realised only by the symmetric theory. Meanwhile, the Spin(8) theory has neither $s$, $\sigma_A$ nor $\rho_B$ fermions. In this case, there is no obstacle to supersymmetrising either of these theories, and the anomaly matching finds a natural home in the supersymmetric Pouliot-Strassler duality \cite{ps1}.

For other values of $p$ the chiral theories -- with both symmetric and anti-symmetric fermions  -- can be supersymmetrised and there is a long literature discussing the resulting dynamics. There is a known dual for the $p=1$ supersymmetric chiral theory, which is based around a Spin(10) gauge group \cite{ps10}. There are also known duals for the supersymmetrised chiral theories involving an anti-symmetric tensor  in which the anti-symmetric chiral multiplet arises as a composite from the confining dynamics of an auxiliary gauge group \cite{micha,poul}.  (See also \cite{hooked,hooke}.) 
Much of the interest in these theories and their duals revolves around the fact that they tend to dynamically break supersymmetry \cite{ads,erich}. However, to our knowledge, the anomaly matching between the chiral theories and the Spin(8) theory described above -- which suggests a natural extension of the Pouliot-Strassler duality \cite{ps1} -- has not been previously noted.  Indeed, as we now explain, 
if there is a duality between these two theories then it is not one that enjoys supersymmetry.

While there is no difficulty in adding scalars to make the chiral $SU(N)$ theories supersymmetric, this is not possible for the Spin(8) theory when $p\neq 0$.
%
At first glance, this seems surprising. After all, the fermion $\tilde{\lambda}$ is nicely placed to play the role of the gaugino, while all other fermions would seem to sit nicely in chiral multiplets.  Nonetheless, naive attempts at making the theory supersymmetric fail.  Here we explain why.

Suppose that $\tilde{\lambda}$ is a gaugino of a supersymmetric Spin(8) theory, with all other fermions sitting in chiral multiplets. We will denote these superfields with hats, so $\hat{v}$ is the chiral superfield containing the fermion $v$, and so on. We would like to write down interaction terms so that the global symmetry of the theory is $G$. This is straightforward for $v$, $s$ and $c$ but, as we've seen above, we must write down further interactions that imprint the correct symmetry transformations on the gauge singlets. One can check that there is a superpotential term that correctly implements the transformation for $\hat{\sigma}_S$ and $\hat{u}$: these are
\be W = \hat{u}\hat{c}\hat{c} + \hat{\sigma}_S\hat{v}\hat{v}\nn\ee
This is neutral under $SU(p)\times SU(q) \times U(1)_1\times U(1)_2$ and has charge +2 under $R_{\rm susy}= R/(N+2)$. Here the normalisation ensures that $\tilde{\lambda}$ has $R_{\rm susy}$ charge $+1$. (And we must remember that $R_{\rm susy} ({\rm boson}) = R_{\rm susy}{(\rm fermion})+1$ for chiral superfields.) However, no further superpotential terms are possible preserving all symmetries, and with only positive powers of fields. For example, one can construct the term $(\hat{s}\hat{v}\hat{c}\hat{\rho}_B\hat{\sigma}_S)^2\hat{u}$, which imposes the right $SU(q)\times SU(p) \times U(1)_1\times U(1)_2$ symmetry structure on $\hat{\rho}_B$, but not the right R-charge. There is no holomorphic term that imposes the right symmetry structure on $\hat{\sigma}_A$. (The choice $\hat{\sigma}_A\hat{\rho}_B\hat{\rho}_B\hat{\sigma}_S$ vanishes on symmetry grounds, and also carries the wrong R-charge.)

One may wonder if we can make progress by introducing yet further fields. These additional fields cannot contribute to the anomalies of $G$, and so should be vector-like. But then we can always give them a mass and integrate them out and any constraints that they impose should be reproducible using just irrelevant operators for the massless fields. And, as we have seen, no such superpotential is possible. It appears that any duality that exists between these theories must be non-supersymmetric\footnote{One may wonder if the symmetry transformations can be imprinted by irrelevant terms arising from a non-trivial K\"ahler potential. Even this appears challenging, simply because the obvious combination $\hat{s}^i\hat{s}^j(\hat{\sigma}_A)_{ij}^\dagger$ vanishes on anti-symmetry grounds.}.


\subsection{Discrete Symmetries}\label{discretesec}

As we've seen, the 't Hooft anomalies for the continuous symmetry group $G$ are shared by the $SU(N)$ and $\spina$ theories. But that alone does not guarantee that the symmetry structure matches. One must worry about discrete aspects of the symmetry. In this section we look more carefully at these discrete quotients.

Before we embark on a full analysis, we can flag up the issue of concern. In the $SU(N)$ chiral theory, all fermions have integer $U(1)$ charges, while in the $\spina$ theory, for certain values of $N$, $p$ and $q$, some fermions have half-integer charges. This, in itself, is not necessarily a worry since only gauge invariant operators should match across the duality. However, one can check that the gauge invariant composite fermion $vsc$ has a half-integer charge under some $U(1) \subset G$ whenever $N$ is odd. This means that there is no possibility of a duality between the $SU(N)$ theory and $\spina$ theory as currently stated, at least for $N$ odd. 

This issue, it turns out, lies at the heart of a mismatch between the discrete symmetry structure of the two theories. There is, however, a simple resolution. The $\spina$ theory has an additional $\Z_2$ symmetry which can be taken to act as
\be \Z_2: c\rightarrow -c\label{z2}\ee
If we promote this to a gauge symmetry, then it projects out the offending composite $vsc$ from the spectrum. As we now show, after making this projection the discrete symmetry structure between the two theories matches.

In the rest of this section, we explain this in more detail. Specifically we will show that, after gauging the $\Z_2$ symmetry \eqn{z2},  the faithfully acting global symmetry group of both theories is (roughly)
\be G = \frac{SU(q)\times SU(p) \times U(1)_1\times U(1)_2\times U(1)_R}{\Z_p\times \Z_q}\label{quotient}\ee
(There are a couple of further flourishes that should be included when $N$, $p$ and $q$ are even that turn the ``roughly" into ``exactly". We describe these below.)

\subsubsection*{Discrete Symmetries for the Chiral Theories}

We start with the chiral theories. 
First we show that there are no additional discrete symmetries. These can arise  if there is a $\Z_k$ subgroup of the anomalous $U(1)$ that changes the $\theta$ angle by $2\pi$. Under a general $U(1)$ transformation,
\eq{\chi\to e^{i\omega_\chi}\chi\ ,\ \psi\to e^{i\omega_\psi}\psi\ ,\ \tilde{\psi}\to e^{i\omega_{\tilde{\psi}}}\tilde{\psi}\ ,\ \la\to e^{i\omega_\la}\la\ ,\nn}
the $\theta$ angle is shifted by
\eq{\delta\theta= (N+2)\omega_\chi+q\omega_\psi+p\omega_{\tilde{\psi}}+2N\omega_\la\ .\nn}
Is there a choice of $\omega_i$ which is not part of the continuous global symmetry such that $\delta\theta=2\pi$? Using the 3 $U(1)$s, we can always set $\omega_{\chi}=\omega_{\tilde{\psi}}=\omega_\la=0$, so we have
\eq{q\omega_\psi=2\pi\ \Rightarrow\ \omega_\psi=\frac{2\pi}{q}\ .\nn}
But this is equivalent to an $SU(q)$ centre transformation. Hence, there are no additional discrete symmetries.

 Next, we will find the faithful symmetry group acting on gauge invariant operators. In order to do that we need to write the most general transformation acting on the fermions and look for non-trivial solutions that leave all the fermions invariant. 
We  analyse the different factors of $G$ in turn. First, we check that the $U(1)^3$ symmetries are freely acting. Then we show that some combination of  $U(1)$ transformations can be undone by a suitable $SU(q)\times SU(p)$ transformation, resulting in the quotient in \eqn{quotient}.

First $U(1)_R$. It is simple to check that this is freely acting if $N$ is odd, but can be quotiented by $\Z_2$ if $N$ is even, and by $\Z_4$ if $N=2\ mod\ 4$. This same behaviour occurs for the Spin(8) theory too. This is a rather trivial quotient and we ignore it in what follows.

Next, we turn to $U(1)_1\times U(1)_2$. For the symmetric theory, we define the linear combination of the $U(1)_1\times U(1)_2$ charges $Q_1$ and $Q_2$
\be Q_A=\onov{N}(Q_2-Q_1)\ \ \ {\rm and}\ \ \  Q_B=\frac{q}{N(N+2)}(Q_1-Q_2)+\frac{1}{N+2}Q_1 \label{AB}\ee
For the anti-symmetric theory, we take the same linear combination but with $N=-\tilde{N}$.  The charges of the fermions $\chi$, $\psi$ and $\tilde{\psi}$ under these redefined $U(1)$'s are given by
\be
\begin{array}{|c||c|c|}
\hline  & U(1)_{A} & U(1)_{B} \\
\hline \rule{0pt}{2.5ex}  
\chi & (N+4)/{N}& -{2q}/{N} \\
\psi &  -(N+2)/{N} & (N+q)/{N} \\
\tilde{\psi} & (N+2)/{N} & -{q}/{N} \\
 \hline
\end{array}
\nn\ee
These are fractional. However, all gauge invariant operators have integer charges. To see this, it suffices to check that a $2\pi$ rotation by either $U(1)_{A}$ or $U(1)_B$  is proportional to the action of the centre of the $SU(N)$ gauge group.
%
%
%
%
Furthermore,  the two gauge invariant operators $\tilde{\psi}\chi^\dagger\tilde{\psi}$ and $\psi\tilde{\psi}$ have  $U(1)_A\times U(1)_B$ charges of $(1,0)$ and $(0,1)$ respectively.  Since there exist gauge invariant operators with the minimal charges, no quotient acting solely on these two $U(1)$s is possible. 

It is, however, possible that some combination of $U(1)$ transformations coincides with the centres of the non-Abelian groups. The transformation of $\lambda$ means that there is no possibility that a $U(1)_R$ can participate in this game, but we must look more carefully at $U(1)_A\times U(1)_B$. 
 We parameterise the centres by  $k,l,m\in \Z$ for $SU(N)$, $SU(q)$ and $SU(p)$ respectively, and denote the $U(1)$ transformations as $e^{2\pi i Q_A\alpha }$ and $e^{2\pi iQ_B\beta}$. The action of these combined symmetries leave $\chi$, $\psi$ and $\tilde{\psi}$ invariant only if the following conditions hold:
\be \chi:\ \ \ &&\frac{2k}{N}+\frac{N+4}{N}\alpha-\frac{2q}{N}\beta\ \in\ \Z \nn\\
	\psi:\ \ \ &&-\frac{k}{N}+\frac{l}{q}-\frac{N+2}{N}\alpha+\frac{N+q}{N}\beta\ \in\ \Z\nn\\
\tilde{\psi}:\ \ \ &&\frac{k}{N}+\frac{m}{p}+\frac{N+2}{N}\alpha-\frac{q}{N}\beta \ \in\ \Z
\nn\ee
There are two solutions. The first is
\be k=-1\ \ ,\ \ l=1\ \ ,\ \ m=0\ \ ,\ \ \alpha=0\ \ ,\ \ \beta=-\onov{q}\label{sol1}\ee
This is the quotient by $\Z_q$ in \eqn{quotient}.  The second is
\be k=-1\ \ ,\ \ l=0\ \ ,\ \ m=1\ \ ,\ \ \alpha=-\frac{2}{p}\ \ ,\ \ \beta=-\onov{p}\label{sol2}\ee
This is the quotient by $\Z_p$ in \eqn{quotient}.

If both $p$ and $q$ are even then there is a $\Z_2\subset\Z_q\times \Z_p$ quotient that acts only on the centres of the non-Abelian groups,  without involving the $U(1)^3$. 
There is one final subtlety. It turns out that some combination of the global symmetry transformation coincides with the fermion number $(-1)^F$.  This $\Z_2$ quotient arises from $k=-1$  $SU(N)$ centre transformation, coupled with a $U(1)_A\times U(1)_R$ transformation which, combined, satisfy
\be  \frac{k}{N} + \left(\frac{1}{2} - \frac{N}{N+2}\right) Q_A + \frac{1}{2(N+2)}{Q_R} = \left\{\begin{array}{cc} \frac{1}{2}\ {\rm mod}\ \Z\ \ \ & \ \ \mbox{for fermions} \\ 0\ {\rm mod}\ Z\ \ & \ \ \mbox{for bosons}\end{array}\right.\label{fermion}\ee
where we should take $k=-1$. This means that the true symmetry group takes the form
\be \frac{Spin(1,3)\times G}{\Z_2}\nn\ee

\subsubsection*{Discrete Symmetries for the Spin(8) Theory}

Now we turn to the Spin(8) theory. For the proposed duality to have a chance of success, the global structure of the symmetry group should agree. We will see that this is indeed the case after  we gauge a $\Z_2$ symmetry of the Spin(8) symmetry.

As with the chiral theories, we will start by checking whether there are additional discrete transformations that change $\theta$ by $2\pi$. Under a general $U(1)$ rotation of the charged fields,
\eq{v\to e^{i\omega_v}v\ ,\ s\to e^{i\omega_s}s\ ,\ c\to e^{i\omega_c}c\ ,\ \tilde{\la}\to e^{i\omega_{\tilde{\la}}}\tilde{\la}\ ,\nn}
the Spin(8)  $\theta$-angle is shifted by
\eq{\delta\theta=12\omega_{\tilde{\la}}+2\omega_c+2p\omega_s+2q\omega_v\ .\nn}
Using the 3 $U(1)$s we can set $\omega_v=\omega_s=\omega_{\tilde{\la}}=0$. We are left with
\eq{2\omega_c=2\pi\ \Rightarrow\ \omega_c=\pi\ .\nn}
We see that the $\Z_2$ transformation \eqn{z2}, which acts as $c\to-c$, is a global symmetry of the theory. In what follows we will find various discrepancies between the global symmetry structure of the $SU(N)$ theory and the Spin(8) theory. All of them are resolved if this $\Z_2$ symmetry is gauged.

First, we ask: is this $\Z_2$ contained inside the other global symmetries? If $N$ is odd, a $U(1)_A$ transformation by $2\pi$ acts as  $(v,s,c)\to (-v,-s,-c)$. If we accompany this by a centre transformation of Spin(8), it  is equivalent to $c\to-c$. Hence, in this case the $\Z_2$ is contained in $U(1)_A$. This is related to the fact, mentioned above, that the gauge invariant operator $vsc$ has half-integer charge when $N$ is odd.  In contrast, when  $N$ is even the $\Z_2$ is not contained within the other symmetry.  To see this, we exhibit gauge invariant operators that are charged only under $c\to-c$. Such an operator is given by  $v^qs^pc\, tr(\tilde{\la}^{ 6})$ if $q,p$ are odd, and $v^{q-1}s^{p+1}c\, \rho_B tr(\tilde{\la}^5)$  if $q,p$ are even. Although the underlying reasons are different, for both $N$ odd and $N$ even we should gauge the $\Z_2$ symmetry \eqn{z2}.

Now we can move on to the quotients.
If $N=0,2\ mod\ 4$ then we should quotient the $U(1)_R$ symmetry by a $\Z_2,\ \Z_4$ respectively to ensure that it acts faithfully. This is the same structure that we saw in the $SU(N)$ theory.

Next, under the linear combinations of charges \eqn{AB} that form $U(1)_A$ and $U(1)_B$, the various fields of the Spin(8) theory transform as
\be
\begin{array}{|c||c|c|}
\hline  & U(1)_{A} & U(1)_{B} \\
\hline \rule{0pt}{2.5ex}  
v & 1/2& -1 \\
s &  -1/2 & 0\\
c & -(N+4)/2 & q\\
\sigma_S & -1 & 2 \\
\sigma_A & -1 & 0\\
\rho_B & 1 & -1 \\
u & N+4 & -2q\\
 \hline
\end{array}
\nn\ee
The adjoint fermion has $A[\tilde{\lambda}] = B[\tilde{\lambda}]=0$. A number of these fields have $1/2$ integer $U(1)_A$ charges. One can check that all gauge invariant fields have integer charges with one exception: the composite $vsc$ has $1/2$ integer $U(1)_A$ charge when $N$ is odd. Hence, as it stands, this field can have no counterpart in the chiral description. As we mentioned previously, this discrepancy is resolved by gauging the $\Z_2$ symmetry \eqn{z2}. This removes the offending states from the spectrum.

After this gauging, the remaining symmetry structure coincides with that of the chiral theory. First, one can check that there are gauge invariant operators with minimal $U(1)_A$ and $U(1)_B$ charges (e.g. $\sigma_A$ and $\sigma_A\rho_B$). Furthermore, one can show that the transformations \eqn{sol1} and \eqn{sol2} act trivially on all fields and so the faithful action of the global symmetry again enjoys the quotient $\Z_p\times \Z_q$ seen in \eqn{quotient}. (Note: one should neglect the $k=-1$ centre of $SU(N)$ when implementing \eqn{sol1} and \eqn{sol2} on the Spin(8) fields.) 

The last remaining subtlety involves the fermion number. The constraint \eqn{fermion} again holds for {\it almost} all gauge invariant operators (this time with $k=0$ as there is no $SU(N)$ gauge symmetry). The ``almost" is because there is one exception: the fermion $vsc$ obeys
\be \left(\frac{1}{2} -\frac{N}{N+2}\right)Q_A[vsc] + \frac{1}{2(N+2)}Q_R[vsc] = \frac{N}{2}\nn\ee
This is inconsistent with the spin relation when $N$ is even. Again, this discrepancy is resolved if we act with the $\Z_2$ quotient \eqn{z2}. 

The upshot of this analysis is that we should gauge the additional $\Z_2$ symmetry both for $N$ even and $N$ odd. The global symmetry structure of the two conjectured dual theories then agrees.

\subsubsection*{There Are No Further Discrete Anomalies}

One may worry that there may be further discrete 't Hooft anomalies that must also be matched between the two theories. In fact, as we now explain, the matching of perturbative 't Hooft anomalies for $SU(q)\times SU(p)\times U(1)^3$ is sufficient to ensure the matching of the faithful symmetry group $G$ given by \eqn{quotient}.

This follows from the following, general result.  Suppose that  $\mathcal{T}_{1}$ and ${\cal T}_2$ are two theories that share the same symmetry group $G= {H}/{\Gamma}$, where $\Gamma$ is a discrete subgroup of $H$. If $\Pi_0(H)=0$, then 't Hooft anomaly matching  for $G$ follows from anomaly matching for $H$.  To understand why, we can define the combined theory $\mathcal{T}_+=\mathcal{T}_1+\mathcal{T}_2^\dagger$. $\mathcal{T}_+$ has no anomalies for $H$. This means that if we turn on background gauge fields for $H$, symmetry transformations don't generate any $\theta$-term for these background fields. We write these as $\bar{\theta}$ to distinguish them from the dynamical theta terms that appeared in our earlier analysis. The lack of anomalies for $H$ means that $\delta\bar{\theta}=0$. The quotient by $\Gamma$ can change the periodicity of $\bar{\theta}$, but this doesn't change the fact that $\delta\bar{\theta}=0$ exactly. Hence, $\mathcal{T}_+$ has no anomalies for $G$.\footnote{The requirement that $\Pi_0(H)=0$ is somewhat hidden in the discussion. This requirement is needed because, even though $\mathcal{T}_+$ has no anomalies for $H$, a non-trivial element of $\Pi_0(H)$ can act as $\delta\bar{\theta}=2\pi$. The quotient by $\Gamma$ can change the periodicity of $\bar{\theta}$ such that $\delta\bar{\theta}=2\pi$ is non-trivial and $\mathcal{T}_+$ has an anomaly for $G$ even when there is no anomaly for $H$.} 

\subsection{Putative Dynamics}

The $SU(N)$ and $\spina$ gauge theories share the same symmetry structure and the same anomalies. The obvious question is: so what? Is there some way in which these theories are dynamically related, presumably for some particular range of $N$, $p$ and $q$?

This is a difficult question to answer. It may be that the two theories are dynamically related, but only after suitable scalars are added, with a well-designed potential coaxing one theory to flow to the other. Supersymmetric theories are, of course, in this category but understanding similarly detailed phenomena in the absence of supersymmetry lies well beyond our current grasp of strongly coupled quantum field theory. Nonetheless, given the anomaly matching it seems reasonable to conjecture that, for a certain range of parameters, the chiral $SU(N)$ gauge theories may flow to the $\spina$ theory at low energies.  In this section and the next, we make a few obvious comments about possible relations between the two theories.

We start with the chiral $SU(N)$ theories. We will consider the theories without additional scalars; these change some of the specific numbers below, but not the overall story. The $SU(N)$ theory with a symmetric fermion is asymptotically free provided:
\be q> q_{AF} = \frac{1}{7}(9p + 34)\nn\ee
(Recall that $q=N+p+4$ so this inequality is telling us that the number of colours $N$ should be suitably larger than the number of flavours $p$, as expected.)
For $q\leq q_{AF}$, the $SU(N)$ symmetric theory is infra-red free. For $q$ sitting slightly above the value $q_{AF}$, the theory flows to an interacting, chiral Banks-Zaks fixed point, with the global symmetry group $G$ unbroken. This conformal behaviour, with unbroken $G$ will persist for some range 
 $q\in (q_{AF},q_{\rm crit})$ that defines the conformal window. As in QCD, the value of $q_{\rm crit}$ that determines the strongly coupled end of the conformal window is unknown.  Our interest here lies in those values $q>q_{\rm crit}$ that are beyond the conformal window. The theory may, of course, break the global symmetry $G$. Alternatively, it may flow to the $\spina$ theory described above.

We have one data point where we do understand quantitatively what happens. This is the case $p=0$ where we add scalars so the theory becomes supersymmetric. Here the chiral $SU(N)$ theory flows to the supersymmetric $\spina$ theory only for $N\geq 13$ or, equivalently, $q\geq 17$ \cite{ps1}. 

If there is indeed a flow from the non-supersymmetric chiral theories to the \spina\ theory, then it should be possible to match gauge invariant operators across the duality. Indeed, there is a fairly natural conjecture for the gauge singlet fermions in the $\spina$ theory. Their quantum numbers coincide with the following gauge-invariant, composite operators of the $SU(N)$ theory,  
\be
\sigma_S \sim \chi\lambda^2\psi^2\ \ \ ,\ \ \ \sigma_A\sim \chi \lambda^2\tilde{\psi}^2\ \ \ ,\ \ \  \rho\sim \chi^\dag(\lambda^\dag)^2\psi^\dag\tilde{\psi}\ \ \ ,\ \ \  u\sim \chi^N\lambda^{N-1}\label{comp}
\ee
where $u\sim \epsilon_{i_1...i_N}\epsilon_{j_1...j_N}\chi^{i_1j_1}...\chi^{i_Nj_N}\text{tr}(\lambda...\lambda)$.

There is a similar story for the $SU(N)$ chiral theory with an anti-symmetric fermion. Now the asymptotic freedom bound is
\be q < \tilde{q}_{AF} = \frac{1}{9}(7p + 34)\nn\ee
Note that the sign of the inequality is reversed. For $q \geq\tilde{q}_{AF}$, the chiral theory is infra-red free. For $q\in (\tilde{q}_{\rm crit},\tilde{q}_{AF})$, the theory flows to an interacting fixed point with $\tilde{q}_{\rm crit}$ unknown. Again, our interest is in the strong coupling dynamics for $q<\tilde{q}_{\rm crit}$ where the $\spina$ theory provides  a putative end point for the flow.

This story is largely consistent with the dynamics of the $\spina$ theory, which is asymptotically free only if
\be q +p< 34\nn\ee
This suggests that there are large swathe of parameter space in which the chiral $SU(N)$ theory may flow to an infra-red free $\spina$ theory.

\begin{figure}[h]
	\centering
	\includegraphics[width=0.5 \linewidth]{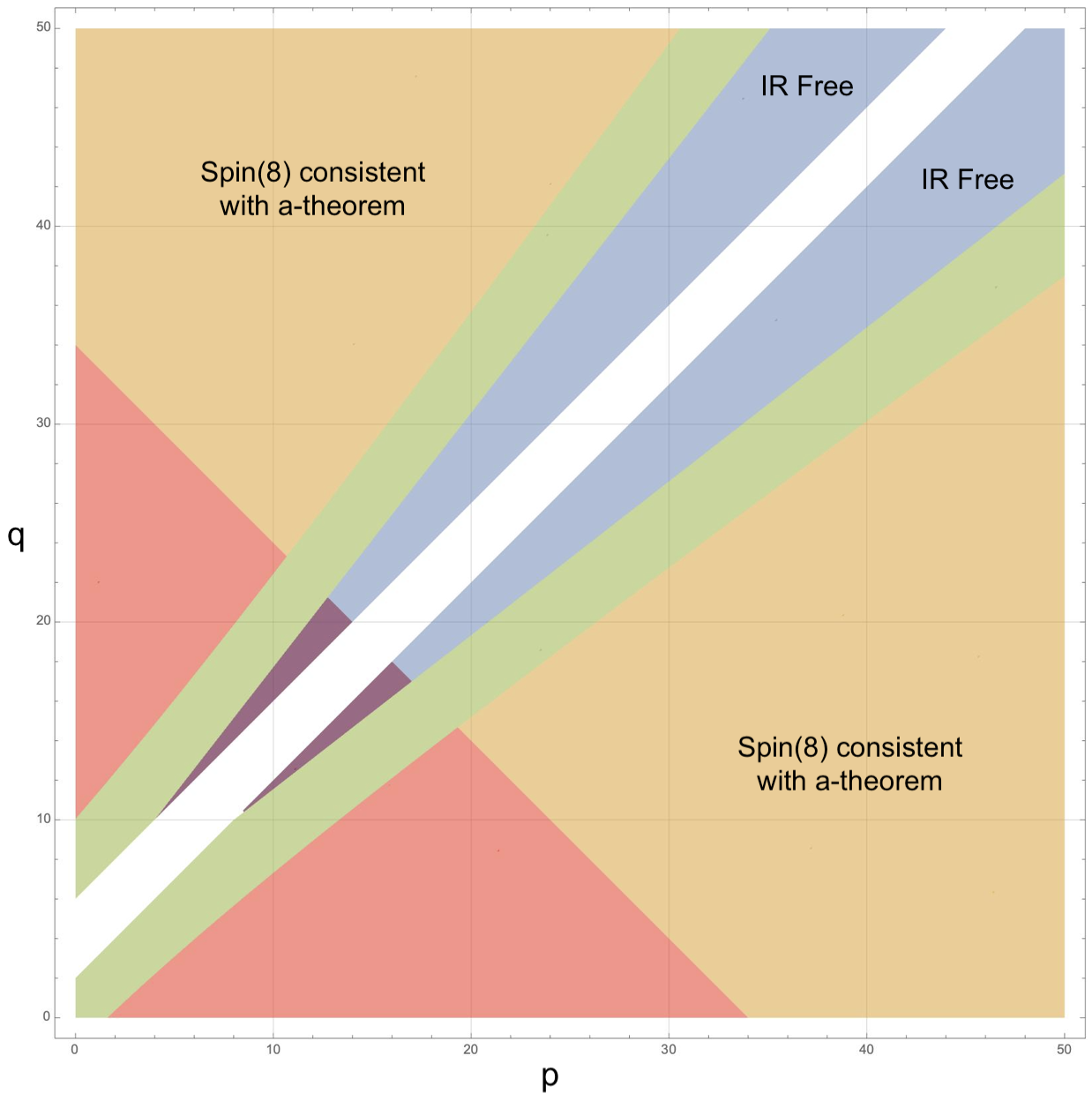}
	\caption{Candidate low energy phases of the chiral $SU(N)$ theories, plotted as a function of $q$ and $p$. The upper region is the chiral theory with a symmetric fermion and the lower with an anti-symmetric fermion. In the white strip, there are no chiral theories carrying a representation of the global symmetry group $G$.  In the blue strip, the $SU(N)$ chiral theory is IR free. The green region is where  flows to the Spin(8) theory are inconsistent and form a potential conformal window for the chiral gauge theory. The orange and light red regions are where the flow from the chiral theories to the $\spina$ theory is consistent with the $a$-theorem; in the orange region the Spin(8) theory is IR free and in the light red region it is asymptotically free. Finally, the dark red region shows where a flow from Spin(8) to the infra-red free chiral theory is feasible.}
	\label{qpfig}
\end{figure}

Further constraints on the allowed RG flows come from the a-theorem. For the symmetric chiral theory, we have  (again, absorbing a factor of
 $90(8\pi)^2$),
\be
a_s=\dfrac{1}{4}\left[303q^2-2325q-562pq+7p(37p+307)+4182\right]
\nn\ee
Meanwhile, for the Spin(8) theory, 
\be
a_8=\dfrac{1}{4}\left[11(p^2+q^2)+187q+11p(2q+15)+8430\right]
\nn\ee
In the large $N$ limit, this tells us that the symmetric chiral theory can flow to the \spina\ theory provided
\be 
N\gtrsim 0.39 p\nn\ee
Similarly the anti-symmetric chiral theory has
\be
a_a=\dfrac{1}{4}\left[303p^2+7q(37q-307)+p(-562q+2325)+4182\right]
\nn\ee
At large $N$, this theory can flow to the \spina\ theory provided that
\be \tilde{N} \gtrsim 0.39 q\nn\ee
These results are plotted in the $(q,p)$-plane in Figure \ref{qpfig}. 

\begin{figure}[h]
	\centering
	\includegraphics[width=0.4 \linewidth]{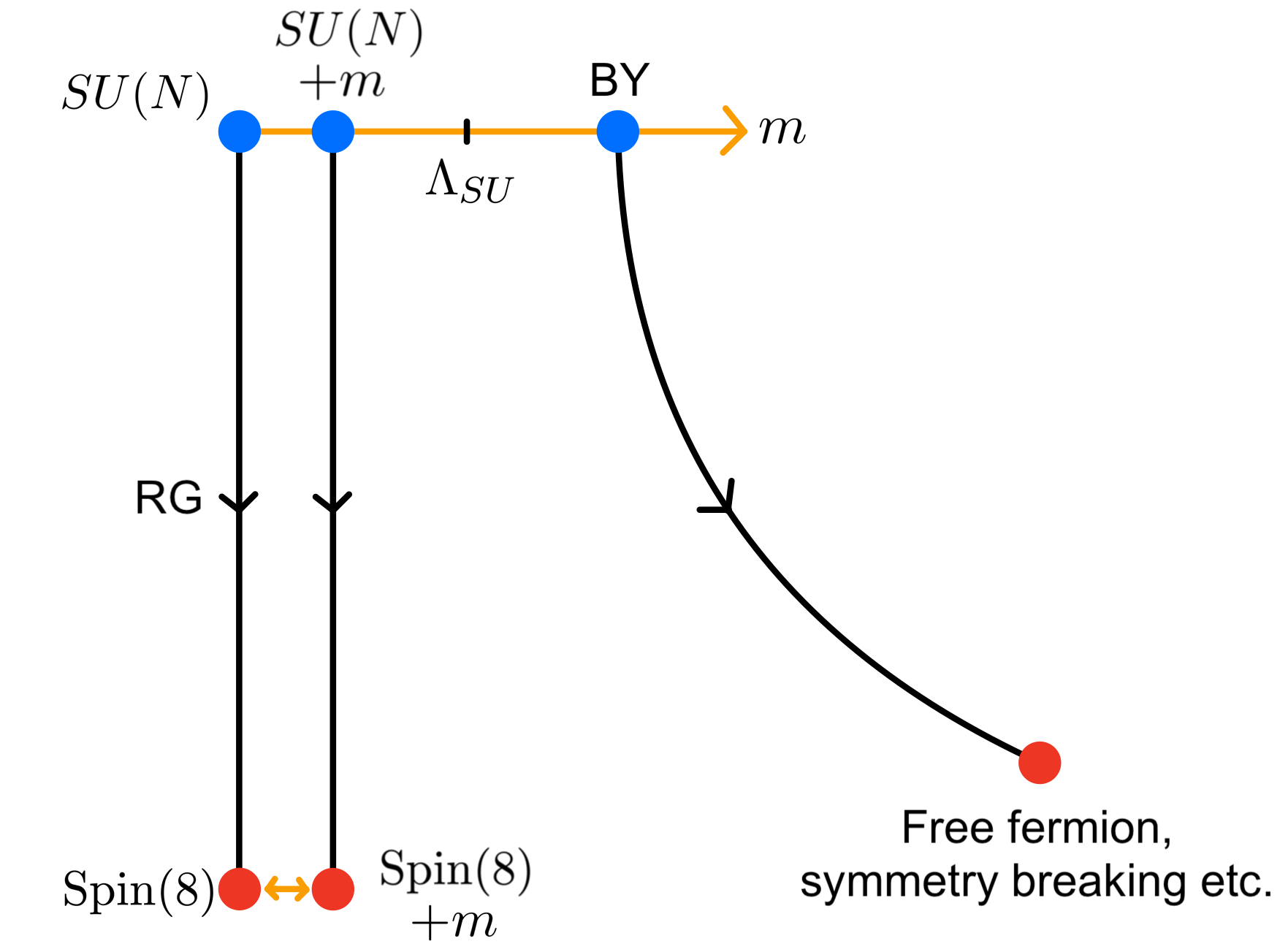}
	\caption{Possible RG flows of the chiral $SU(N)$ theories as they are deformed by a mass term for the adjoint fermion.}
	\label{rgfig}
\end{figure}

It has been known since the initial work of  Pouliot \cite{ps0} that chiral gauge theories may have non-chiral duals. In the known supersymmetric examples, a chiral theory in the UV always flows to a non-chiral theory in the IR. It is interesting to ask whether the inverse flow is possible, with chirality -- in the sense that no gauge invariant mass terms are allowed -- emerging in the IR. There is nothing to prohibit such a flow in principle, and it would be interesting as it may offer an alternative avenue to putting chiral gauge theories on the lattice.  Taken at face value, the results above suggest that there is a small window, shown by the red region in Figure \ref{qpfig},  in which the $\spina$ theory is asymptotically free and may flow to a (admittedly weakly coupled) chiral $SU(N)$ theory. The window is not parametrically large and it quickly closes if one starts to introduce further scalars. Nonetheless, this behaviour remains as an interesting possibility.

\subsubsection*{Mass Deformations}

If the non-supersymmetric flow from the chiral theory to the \spina\ theory does take place for some range of parameters, then it has an interesting consequence. Suppose we deform the chiral theory by giving a mass to the adjoint fermion $\lambda$,
\be
\delta \mathcal{L}= m\,{\rm tr}\, \lambda \lambda + {\rm h.c.} \nn
\ee
 From \eqn{syml} or \eqn{antil}, we see that this breaks the $U(1)_R$ symmetry, leaving the other global symmetry groups intact.  What is the effect of this mass deformation in the infra-red?

For $m\ll \Lambda$, the strong coupling scale of the chiral gauge theory, it should be possible to see the effect directly in the low-energy effective theory. If this is the \spina\ dual, then the deformation naturally maps to a mass term for the adjoint fermion $\tilde{\lambda}$, again breaking the $U(1)_R$ symmetry. No other relevant deformations are possible without further breaking the global symmetry.

In particular, the gauge singlets  $\sigma_S$, $\sigma_A$, $\rho_B$ and $u$ must all remain massless after this deformation. Yet, as shown in \eqn{comp}, each is a composite operator that contains the now-massive fermion $\lambda$. If the duality holds there must be massless composite fermions that contain a massive constituent. This behaviour is surprising and is in violation of the so-called ``persistence of mass condition" \cite{preskill}. (The fact that non-supersymmetric dualities violate the persistence of mass condition was previously noted in \cite{ofer}.) Indeed, such behaviour is ruled out in QCD-like theories by the Vafa-Witten theorem \cite{vw}. But the Vafa-Witten theorem relies on the positive definiteness of the fermionic measure, which is true in QCD but not in the chiral gauge theories that are of interest here.

It is unclear what to make of this. Depending on taste, this behaviour could be viewed as a hint that the duality does not, in fact, hold, or as a hint that strongly interacting chiral gauge theories can do interesting things.

In the limit  $m\gg \Lambda$, we can integrate out the adjoint fermion and recover the Bars-Yankielowicz chiral theories of Section \ref{bysec}. There does not appear to be any relevant operator that flows from the \spina\ theory to the free fermions \eqn{byfreefermion}. This means that there should be a phase transition as the mass is increased. This is shown schematically in Figure \ref{rgfig}. Presumably, at the separatrix between the different end points, there should be a novel fixed point.

As an aside: one might naively think that one could integrate out $\tilde{\la}$ to be left with a Spin(8) theory without an adjoint fermion that shares the same 't Hooft anomalies as the $SU(N)$ chiral gauge theories that we met in Section \ref{bysec}. While it is true that the perturbative 't Hooft anomalies agree, the global structure of the symmetry group does not. For this reasons, we  do not know of a Spin(8) dual for the Bars-Yankielowicz theories.

\section{Deformations of Supersymmetric Dualities}\label{4sec}

In this final section, we review and slightly extend a number of interesting comments and proposals for non-supersymmetric dualities that can be found in the literature.

\subsection{Vector-Like Dualities Without Scalars}\label{vlduals}

One  way to search for non-supersymmetric dualities is to start from well known supersymmetric dualities and softly break supersymmetry on both sides. There have been many attempts at doing so, including \cite{ofer, yael,ah,mury}. But retaining control in the strong coupling regime after supersymmetry breaking is challenging and, to date, there are no well-established dualities beyond those that arise trivially through soft supersymmetry breaking. 

One may try to be bolder. The 't Hooft anomaly matching conditions that hold for Seiberg dualities also trivially hold for the theories in which all scalar superpartners are simply discarded. For example, from the original Seiberg duality \cite{seib2}, the QCD-like theory with the following fermions
\be
{\rm Electric\ Theory:}&& 
\begin{array}{|c||c|c|c|c|c|}
\hline & SU(N_c) & SU(N_f)_L & SU(N_f)_R & U(1)_B & U(1)_R \\
\hline \lambda & \operatorname{adj} & \mathbf{1} & \mathbf{1}  & 0 & N_f \\
\psi & \Box & \Box & \mathbf{1} & +1 & -N_c  \\
\tilde{\psi} & \overline{\Box} & \mathbf{1}& \overline{\Box} & -1 & -N_c  \\
\hline
\end{array}
\nn\ee
has the same 't Hooft anomalies as the following fermions of the magnetic theory
\be
{\rm Magnetic\ Theory:}&&
\begin{array}{|c||c|c|c|c|c|}
\hline & SU(N_f-N_c) & SU(N_f)_L & SU(N_f)_R & U(1)_B & U(1)_R \\
\hline \tilde{\lambda} & \operatorname{adj} & \mathbf{1} & \mathbf{1}  & 0 & N_f \\
q & \Box & \overline{\Box} & \mathbf{1} & +1 & N_c-N_f  \\
\tilde{q} & \overline{\Box} & \mathbf{1} & \Box  & -1 & N_c-N_f  \\
\mu & \mathbf{1} & \Box & \overline{\Box} & 0 & N_f-2N_c\\
\hline
\end{array}
\nn\ee
Might it be that, for some values of $N_c$ and $N_f\geq N_c$, the non-supersymmetric electric theory flows to the magnetic theory, preserving the $SU(N_f)_L\times SU(N_f)_R\times U(1)_B\times U(1)_R$ global symmetry?

The answer is an unequivocal no \cite{ofer}. This follows from various QCD theorems. The key observation is that the determinant that arises from integrating out the fermions is positive definite. (This is not true of the chiral theories described in earlier section, and nor is it true of supersymmetric Seiberg dualities where the presence of scalars with Yukawa couplings is sufficient to ensure that the fermion determinant can have both positive and negative sign.) This fact allows us to invoke various  Vafa-Witten-Weingarten theorems.

There are two such arguments \cite{ofer}.  The first uses the Weingarten inequality \cite{wein}. Roughly speaking, in the context of QCD, the Weingarten inequality states that the proton is always heavier than the pion. Crucially, the proof of the inequality assumes neither confinement nor chiral symmetry breaking. In the present context, the Weingarten inequality can be applied to the gauge invariant singlet in the infa-red. This is a composite fermion, with UV constituents given by
\be \mu = \tilde{\psi}\lambda\psi \label{seibergferm}\ee
The Weingarten inequality can be used to show that $\mu$ is always heavier than the scalar formed from  $\tilde{\psi}\psi$. But if all global symmetries are unbroken, and in the absence of fine-tuning, then the scalar $\tilde{\psi}\psi$ is surely massive. The Weingarten inequality then states that $\mu$ must also be massive and so the 't Hooft anomalies cannot be saturated by the matching above. 

The Weingarten inequalities also apply, with some small amendments, to the $SO(N)$ and $Sp(N)$ dualities of \cite{is} and \cite{ip} in the absence of scalars. We describe this in some detail in  Appendix \ref{weinapp}.

The second argument of \cite{ofer} arises if we deform the electric theory by providing a small mass for the adjoint fermion $\lambda$. This breaks the $U(1)_R$ symmetry, leaving the remaining $SU(N_f)_L\times SU(N_f)_R\times U(1)_B$ untouched. In the infra-red, magnetic theory the only such deformation consistent with these symmetries is a mass for the adjoint fermion $\tilde{\lambda}$. In particular, the gauge singlet $\mu$ must remain massless. But $\mu$ is a composite fermion, given in terms of UV constituents as \eqref{seibergferm} and includes the now massive $\lambda$. The proposal that $\mu$ remains massless after the deformation is therefore in contradiction with the Vafa-Witten theorem of \cite{vw}. This states  that -- under the assumption that the fermion determinant is positive definite --  a massless, composite fermion cannot contain a massive constituent. 

We note that both of these objections disappear if a gauge singlet $\tilde{\mu}$, with quantum numbers conjugate to $\mu$, is added to the electric theory which is then deformed by a four fermion term ${\cal L}\sim \tilde{\mu} \tilde{\psi}\lambda\psi$. For values of $N_c$ and $N_f$ for which this is dangerously irrelevant, it is plausible that the theory flows to the magnetic theory, now without the gauge singlet $\mu$.

\subsection{Non-Supersymmetric $Sp(N)$ and $SO(N)$ Dualities}\label{spsec}

A number of proposals for non-supersymmetric dualities exist in the literature, inspired by various orientifold constructions in string theory \cite{moreadi,hook}. In particular, motivated by D-branes, Armoni suggested a non-supersymmetric modification of Seiberg duality for $Sp(N)$ and $SO(N)$ gauge groups \cite{adi1,adi2}. Here we make some comments on this proposal. We first describe the $Sp(N)$ duality, and then the $SO(N)$ duality.

A well-established duality for supersymmetric $Sp(N)$ gauge theories was discovered by Intriligator and Pouliot \cite{ip}.  For $Sp(N)$ gauge group, the adjoint representation coincides with the $\sym$. The proposal of Armoni is to exchange the gaugino in the $\sym$ representation for a Weyl fermion in the $\anti$ representation. The resulting theories have global symmetry
\be G = SU(2N_f)\times U(1)\nn\ee
The electric theory has fermionic matter content
\be
\mbox{Electric Theory:}\ \ \ \ \begin{array}{|c||c|c|c|}
\hline 
& Sp(N_c) & SU(2N_f) & U(1) \\
\hline \rule{0pt}{3.4ex}  \lambda & {\tiny{{\yng(1,1)}}} & \mathbf{1} & -N_f \\
\psi & {\tiny{{\yng(1)}}} & {\tiny{{\yng(1)}}} & N_c-1 \\
\hline
\end{array}\nn
\ee
For $N_f> N_c-2$, the 't Hooft anomalies for $G$ are matched by a putative dual, magnetic theory
\be
\mbox{Magnetic Theory:}\ \ \ \ 
\begin{array}{|c||c|c|c|}
\hline & Sp({N_f-N_c+2}) & SU(2N_f) & U(1) \\
\hline \rule{0pt}{3.4ex}   \tilde{\lambda} & {\tiny{{\yng(1,1)}}} & \mathbf{1} & -N_f \\
\tilde{\psi} & {\tiny{{\yng(1)}}} & \overline{{\tiny{{\yng(1)}}}} & N_f-N_c+1 \\
\mu & \mathbf{1} & {\tiny{{\yng(2)}}} & 2N_c-2-N_f\\
\hline
\end{array}
\nn\ee
When $N_f=N_c-2$, the 't Hooft anomalies are matched by the free fermions $\mu$ and $(\psi\lambda)^\dag \psi$ in the $\mathbf{1}_{(N_c-1)}$ of $G$.

There is a similar story for $SO(N)$ gauge theories. The supersymmetric duality of Intriligator and Seiberg \cite{is} can be deformed by replacing the adjoint gaugino, which sits in the $\anti$ representation, with a fermion in the $\sym$. Now the global symmetry group is
\be  G = SU(N_f) \times U(1)\nn\ee
This time, the electric theory has fermionic matter content
\be
\mbox{Electric Theory:}\ \ \ \ \begin{array}{|c||c|c|c|}
\hline & SO(N_c) & SU(N_f) & U(1) \\
\hline \lambda & {\tiny{{\yng(2)}}} & \mathbf{1} & -N_f \\
\psi & {\tiny{{\yng(1)}}} & {\tiny{{\yng(1)}}} & N_c+2 \\
\hline
\end{array}
\nn\ee
For $N_f > N_c+4$, there is a second, magnetic theory whose 't Hooft anomalies match those of $G$,
\be
\mbox{Magnetic Theory:}\ \ \ \ \begin{array}{|c||c|c|c|}
\hline & SO(N_f-N_c-4) & SU(N_f) & U(1) \\
\hline \rule{0pt}{2.5ex}  \tilde{\lambda} & {\tiny{{\yng(2)}}} & \mathbf{1} & -N_f \\
\tilde{\psi} & {\tiny{{\yng(1)}}} & \overline{{\tiny{{\yng(1)}}}} & N_f-N_c-2 \\
\mu & \mathbf{1} & {\tiny{{\yng(1,1)}}} & 2N_c-N_f+4\\
\hline
\end{array}
\nn\ee
Under what circumstances might the electric theories above flow to the magnetic theories (or, indeed, vice versa)? First, the Vafa-Witten-Weingarten theorems described in the previous section and in more detail in  Appendix \ref{weinapp} ensure that the purely fermionic  electric theory cannot flow to the magnetic theory. We should, therefore, add scalars to the electric theory. 

It is natural to take these, as in the supersymmetric theories, to sit in the representation $(\Box, \Box)_q$, with $q$ the same $U(1)$ charge as $\psi$. We then add additional terms to the action
\be {\cal L} \sim \phi^\dagger \lambda \psi + {\rm h.c.} - V(\phi)\nn\ee
where we are free to tune the potential $V(\phi)$ in the hope of coaxing the electric theory to flow to the magnetic theory. 

As an aside: motivated by the D-brane set-up, Armoni set $V(\phi)=0$, relying on the induced Coleman-Weinberg potential. This immediately rules out the $SO(N)$ duals because the resulting potential pushes $\phi\ra \infty$ on the pseudo-moduli space. This is because bosons give rise to an attractive potential while fermions give rise to a repulsive potential, but for the $SO(N)$ theory there are more fermionic degrees of freedom in the $\sym$ than there are gauge boson degrees of freedom in the adjoint $\anti$. For the $Sp(N)$ theory, it's the other way around and the Coleman-Weinberg potential pushes the theory to the strongly coupled origin of the moduli space where interesting things may happen. However, in the absence of supersymmetry there's nothing to stop us exploring all possible potentials $V(\phi)$.

As always, the question that we would like to ask is: what is the low-energy dynamics of the theories? As usual, this is a difficult question. If the global symmetry group $G$ is unbroken in the infra-red and the magnetic theory is infra-red free, then the magnetic theory provides a good candidate for the low-energy physics. 
There is certainly no contradiction with the a-theorem which, assuming that there are no light scalars, allows a flow from the electric theory to magnetic theory whenever $N_f \lesssim 1.9 N_c$ for both $Sp(N_c)$ and $SO(N_c)$ theories in the largeish $N_c$ limit. 

\appendix

\section{Appendix: Discrete Quotients}\label{quotapp}

In Section \ref{byspecsec}, we described a slight variant on the Bars-Yankielowicz (BY) theories that resulted in a factorised symmetry \eqn{factor}
\be 
G = G_1 \times G_2  = [SU(q)\times U(1)_1'] \times [SU(p)\times U(1)_2']\label{factapp}\ee
Here the factorisation is the statement that the theory has no mixed 't Hooft anomaly between $G_1$ and $G_2$. This observation allowed for a speculative, but plausible, duality proposal.

In this appendix, we point out that the factorisation \eqn{factapp} is not quite complete. There are discrete quotients that relate $G_1$ and $G_2$. We describe these discrete quotients and their implications for the speculative duality of Section \ref{byspecsec}. 

The analysis is almost identical to that of Section \ref{discretesec} where we studied the discrete structure of the theories with an additional adjoint $\la$. Without $\la$ we don't have $U(1)_R$, but otherwise the story is the same. As in Section \ref{discretesec}, it will be convenient to define $U(1)_{A}$ and $U(1)_B$  as in \eqref{AB}, under which the fermions have charges
\be
\begin{array}{|c||c|c|}
	\hline  & U(1)_{A} & U(1)_{B} \\
	\hline \rule{0pt}{2.5ex}  
	\chi & (N+4)/{N}& -{2q}/{N} \\
	\psi &  -(N+2)/{N} & (N+q)/{N} \\
	\tilde{\psi} & (N+2)/{N} & -{q}/{N} \\
	\hline
\end{array}
\nn\ee
These two $U(1)$s act faithfully on gauge invariant operators. The factorisation of the global symmetry group \eqn{factapp} is realised only by defining linear combinations of $U(1)_1'$ and $U(1)_2'$; these are related to $U(1)_A$ and $U(1)_B$ as
\be 
Q_1'=NQ_B\ \ \ {\rm and}\ \ \ Q_2'=2NQ_A+NQ_B\label{ABtoprime}\ee
The centres of the two non-abelian global symmetries $\Z_{q,p}\subset SU(q),\ SU(p)$ are equivalent to a certain $U(1)_A\times U(1)_B$ transformations as described in equations \eqref{sol1} and \eqref{sol2}. In terms of $U(1)'_1$ and $U(1)'_2$ this works as follows: A $\Z_q\subset SU(q)$ is gauge equivalent to $e^{2\pi i Q_1' \omega_1'+2\pi i Q_2'\omega_2'}$ with $\omega_1'=\onov{qN}\ ,\ \omega_2'=0$, and a $\Z_p\subset SU(p)$ is gauge equivalent to $e^{2\pi i Q_1' \omega_1'+2\pi i Q_2'\omega_2'}$ with $\omega_1'=0\ ,\ \omega_2'=\onov{Np}$. If $q,p$ are both even, the $\Z_2\subset \Z_q\times \Z_p$ is gauge equivalent to the identity without the need of using $U(1)$ transformations. 

The last point we want to emphasize is the mixing with fermion number. It is straight forward to show that $e^{2\pi i Q_1' \omega_1'+2\pi i Q_2'\omega_2'}$ with $\omega_1'=-\omega_2'=\onov{4N}$ is equivalent to fermion number.
What does it tell us about the factorized spectrum? Assuming all the operators are charged only under one of the $G_1\times G_2$ factors, the discrete structure puts some constraints on the allowed spectrum. For example, if $p,q$ are both even, the $\Z_2\subset \Z_q\times \Z_p$ implies there are no fundamentals of $SU(q)$ or $SU(p)$ (more generally, no representations with odd number of indices are allowed). The relation \eqref{ABtoprime} implies that all the $U(1)'_{1}$ and $U(1)'_2$ charges are multiples of $2N$ which is not true for general operators charged under both $U(1)$s (recall that $U(1)_{AB}$ and $U(1)_B$ act faithfully). Happily, the mixing with fermion numbers implies that every operator charged only under one of  $U(1)'_{1}$ and $U(1)'_2$ must have charge $2N(1+2\Z)$ for fermions, and $(4N\Z)$ for bosons. So the two requirements are consistent with each other and allow for factorized operators of any spin to exist.

\section{Appendix: Weingarten Inequalities}\label{weinapp}

In this Appendix, we describe in some detail the application of the Weingarten inequality \cite{wein} to rule out certain non-supersymmetric dualities in vector-like theories, as first proposed in \cite{ofer}. We also provide generalisations to $Sp(N)$ and $SO(N)$ dualities, which require small tweaks to the arguments.

\subsubsection*{When Is the Fermion  Determinant Positive Definite?}

The key assumption underlying both Vafa-Witten and Weingarten theorems is that the fermion determinant is positive definite. We first review the arguments for when, and why, this occurs.

The most common setting for the Weingarten inequalities is in QCD, where we are dealing with Dirac fermions $\Psi$.  Consider the Euclidean partition function for a massless Dirac fermion in $d=4$ coupled to gauge fields in an arbitrary representation,
\be
\mathcal{Z}=\int \mathcal{D}A {\cal D}\psi{\cal D}\bar{\psi}\ e^{-S_{\text{YM}}-S_{\text{Fermion}}}=\int \mathcal{D}A\,\operatorname{det}(\slashed{D})\,e^{-S_{\text{YM}}}
\nn\ee
The Dirac operator $i\slashed{D}$ is Hermitian and so its eigenvalues are real 
\be
i\slashed{D} \psi_n =\lambda_n \psi_n \;\;\;\; \lambda_n\in \mathbb{R}
\label{eq:Diraceigenvalues}
\nn\ee
Because $\{\gamma^5, \slashed{D} \}=0$, we also have  $i\slashed{D} (\gamma^5 \psi_n)= -\lambda_n (\gamma^5 \psi_n)$.This means that all non-zero eigenvalues come in plus-minus pairs. There may also be some number, $m$, of zero modes $\tilde{\lambda}$. The general form
of the Dirac determinant is then
\be
\operatorname{det}(\slashed{D})=\tilde{\lambda}^m\prod_{n}(i\lambda_n)=\tilde{\lambda}^m\prod_{\lambda_n>0}(i\lambda_n)(-i\lambda_n)=\tilde{\lambda}^m\prod_{\lambda_n>0} (\lambda_n)^2\geq0
\label{eq:Diracdeterminant}
\ee
which is manifestly positive semi-definite. 

If we couple scalars through a Yukawa coupling, then the Dirac determinant becomes complex.  This means that the Weingarten theorem (or, indeed, the Vafa-Witten theorem) does not apply in the supersymmetric context. This is fortunate because, as we will see, these theorems can be used to rule out certain dualities \cite{ofer}.

What about Weyl fermions that are not in Dirac pairs? Under what circumstances is the path integral measure non-negative? 

To answer this, we first return to a Dirac fermion in $\Psi$ in Lorentzian signature, coupled to a background gauge fields $A$ in some complex representation $R$ of the gauge group $G$. We can write $\Psi$ as two left handed Weyl fermions $\psi$ and $\tilde{\psi}$ in conjugate representations of $G$. Then the partition functions are related by $\mathcal{Z}_{\Psi}=\mathcal{Z}_{\psi}\mathcal{Z}_{\tilde{\psi}}$. Furthermore,  $\mathcal{Z}_{\tilde{\psi}}=\mathcal{Z}_{\psi}^*$ so we can write $\mathcal{Z}_{\Psi}=|\mathcal{Z}_{\psi} |^2$. (See, for example, \cite{wittenfermion}, for a careful study of fermion determinants.) The path integral for a single Weyl fermion then takes the form
\be
\mathcal{Z}_\psi[A]=e^{iW[A]}| \mathcal{Z}_\Psi[A] | = e^{iW[A]} \operatorname{det}(\slashed{D})^{1/2}
\nn
\ee
where we take the positive square-root of the Dirac determinant. It was shown in \cite{ag} that the phase $W[A]$ is given by
\be
W[A]-W[A_0]=\pi \eta[H_5]+2\pi Q_5[A_t]
\nn\ee
where $H_5$ is a five dimensional Dirac operator and $A_t=(1-t)A_0+A$, $t\in [0,1]$ linearly interpolates between a reference configuration $A_0$ to $A$. Here $Q_5[A_t]$ is the Chern-Simons form and encodes chiral anomalies. 

If the representation $R$ of the fermion is complex, then the partition function will in general carry a complex phase. This is a feature of chiral gauge theories and means that  Vafa-Witten-Weingarten theorems do not apply to chiral gauge theories.

In contrast, if we have  Weyl fermions in a real or pseudo-real representation of $G$, then $\mathcal{Z}_{\tilde{\psi}}=\mathcal{Z}_\psi$ and $\mathcal{Z}_{\Psi}=\mathcal{Z}_{\psi}^2$ such that $\mathcal{Z}_{\psi}=\pm\operatorname{det}(\slashed{D})^{1/2}$. We see that the  path integral is real, but it is not necessarily positive semi-definite.

To proceed, we pick a reference gauge configuration $A_0$ and  define the sign to be positive $\mathcal{Z}_{\psi}[A_0]\geq0$. This corresponds to taking the product of the positive eigenvalues of $i\slashed{D}$.  We then ask how the sign varies as we deform $A_0$ to physically distinct gauge configurations $A$. If the sign turns negative for any $A$, the partition function is not positive semi-definite. As we now explain, the partition function obeys $\mathcal{Z}_{\psi}[A]\geq0$ for all gauge configurations $A$ for a Weyl fermion in a real representation of $G$\footnote{We thanks Nakarin Lohitsiri and Tin Sulejmanpasic for explaining to us the importance of Kramers' doubling in the argument below.}. This is not necessarily the case for pseudo-real representations. 

To this end, firstly recall that in $d=4$, we can write a Weyl fermion in a real representation of $G$ as a Dirac-Majorana fermion $\Psi$ with the action
\be
S=i \int d^4x \Psi^T C \slashed{D}\Psi
\nn\ee
where $C$ is the unitary, anti-symmetric charge conjugation matrix satisfying $C\gamma^\mu C^{-1}=-(\gamma^\mu)^T$ and $D_\mu^*=D_\mu$ as $\Psi$ is in a real representation. The partition function formally evaluates to the Pfaffian of the anti-symmetric bilinear form $C\slashed{D}$
\be
\mathcal{Z}[A]=\operatorname{Pf}[C\slashed{D}]
\label{eq:Majoranapfaffian}
\ee
\eqref{eq:Diraceigenvalues} still holds for the Dirac operator $i\slashed{D}$ but in addition we have an extra degeneracy. We can choose the Euclidean gamma matrices $(\gamma^\mu)^\dag =\gamma^\mu$ such that for real representations $C\slashed{D}C^{-1}=-\slashed{D}^*$. Then conjugating \eqref{eq:Diraceigenvalues} gives
\be
iC\slashed{D}C^{-1}\psi_n^*=\lambda_n\psi_n^* \;\;\; \Rightarrow \;\;\; i\slashed{D}(C^{-1}\psi^*_n)=\lambda_n (C^{-1}\psi^*_n)
\label{eq:Kramers_doubling}
\ee
We see that for every eigenvector $\psi_n$, $C^{-1}\psi^*_n$ is also an eigenvector with the same eigenvalue. As discussed in \cite{wittenfermion}, in Euclidean signature, there is no field that corresponds the the complex conjugate of $\psi_n$. However, once we formally write the partition function as a Pfaffian \eqref{eq:Majoranapfaffian}, it makes sense to consider the behaviour of the eigenmodes of the Dirac operator under complex conjugation. Then $\psi_n$ and $C^{-1}\psi^*_n$ are orthogonal to each other due to the anti-symmetry of $C$.

Next, we define the anti-linear operator $\mathcal{T}=C^{-1}K$ with $K$ the complex conjugation operator, we have $[\mathcal{T},i\slashed{D}]=0$ and $\mathcal{T}^2=C^{-1}KC^{-1}K=C^{-1}C^T=-C^{-1}C=-1$. Therefore we can see this as a Kramers degeneracy \cite{wittenfermion}.

Now if we chose the basis $(\psi_1,C^{-1}\psi^*_1,...,\psi_n,C^{-1}\psi^*_n,...)$, the operator $C\slashed{D}$ takes the form
\be
C\slashed{D} = \bigoplus_{n}\left(\begin{array}{cc}
0 & i\lambda_{n} \\
-i\lambda_{n} & 0
\end{array}\right)
\nn\ee
and the partition function \eqref{eq:Majoranapfaffian} evaluates to
\be
\mathcal{Z}[A]={\prod_{n}}'(i\lambda_n)
\nn\ee
where by ${\prod_{n}}'$ we mean in the product we take only one eigenvalue from each Kramers doublet pair. The doubling argument we used in \eqref{eq:Diracdeterminant} by anti-commuting $\gamma^5$ also still holds so we can write the partition function as
\be
\mathcal{Z}= {\prod_{\lambda_n\geq0}}' (\lambda_n)^2
\nn\ee
which is manifestly positive semi-definite for all $A$ and is unaffected by a change of basis as discussed in \cite{stone}. 

The Kramers doubling argument in \eqref{eq:Kramers_doubling} fails for pseudo-real representations and the partition function of these theories are not positive semi-definite. This of course had to be the case in order to be consistent with the Witten anomaly \cite{wittensu2} where the partition function changes sign under a gauge transformation.

\subsubsection*{Weingarten Inequalities for Putative  $Sp(N_c)$ Duals}

We move onto consider a Weingarten analysis of the purely fermionic electric theories introduced in Section \ref{spsec}. We focus on the $Sp(N_c)$ theory as this case is slightly more subtle. The same results trivially apply to the $SO(N_c)$ theories.

Our aim is to show that the gauge singlet composite fermion $\mu \sim \psi \chi \psi$ must be heavier than the meson $ \pi\sim \psi \psi$. For this, we firstly need to make sure that the fermionic path integral is positive definite: this reads
\be
\mathcal{Z}_\text{fermion}[A]= \operatorname{det}(\slashed{D}_\lambda)^{1/2}\operatorname{det}(\slashed{D}_\psi)^{N_f}\nn
\ee
The anti-symmetric representation of $Sp(N_c)$ is real so the discussion above shows $\mathcal{Z}_\text{fermion}[A]\geq 0$ for all gauge configurations $A$. 

The inequalities involving the $Sp(N_c)$ theory are slightly more subtle than the $SU(N_c)$ and $SO(N_c)$ cases. This is simply because the colour indices in $\rho$ and $\pi$ are contracted via $J^{ab}$, the invariant anti-symmetric tensor of $Sp(N_c)$ as opposed to the identity matrix in the $SU(N_c)$ and $SO(N_c)$ cases. We will however see that this does not invalidate the inequalities.

It is convenient to work with Majorana fermions rather than Weyl fermions, so we take the Majorana fermions $\Lambda$ and $N_f$ flavours of $\Psi^i$, which in the Weyl basis is
\be
\Lambda_{ab} = \begin{pmatrix}
        \lambda_{\alpha,ab} \\
        J_{ac}J_{bd}\bar{\lambda}^{\dot{\alpha},cd}
    \end{pmatrix}, \;\;\;\;
\Psi_a = \begin{pmatrix}
        \psi_{\alpha,a} \\
        J_{ab}\bar{\psi}^{\dot{\alpha}b}
    \end{pmatrix}
\label{eq:majweyl}
\ee
 with explicit colour and spinor indices. We are interested in the correlation functions of the composite fermion operator
\be
\mu^{ij}_\alpha =J^{ab}J^{cd}\Gamma^{\beta\gamma\delta}_\alpha \Psi^i_{\beta, a} \Lambda_{\gamma,bc} \Psi^j_{\delta, d}\nn
\ee
where $\Gamma$ is a tensor acting on the spinor indices which we may chose to symmetrise over the flavour indices and project appropriately onto the left-handed spinors. For simplicity, take the flavour indices $i\neq j$ (when they are equal we simply get and extra factor from Wicks theorem) and consider the Euclidean propagator (with no $i$, $j$ sum)
\be
\begin{split}
\langle \mu_\alpha^{ij}(x) \bar{\mu}^{\alpha'}_{ij}(y) \rangle =& -\Gamma^{\beta\gamma\delta}_\alpha \overline{\Gamma}_{\beta'\gamma'\delta'}^{\alpha'} J^{ab}J^{cd}J^{a'b'}J^{c'd'} \\
&\int d\mu \left(S_\Psi(x,y)^{\beta'}_\beta\right)_{aa'} \left(S_\Lambda(x,y)^{\gamma'}_\gamma\right)_{bc,b'c'} \left(S_\Psi(x,y)^{\delta'}_\delta\right)_{dd'}
\nn
\end{split}
\ee
Here we have integrated out the fermions such that $S$ are the fermion propagators in the presence of a fixed background gauge field and $d\mu$ is the path integral measure including the fermion determinants. We take the Frobenius norm over the spinor indices and bound the correlator from above
\be
\begin{split}
|\langle \mu^{ij}(x) \bar{\mu}_{ij}(y) \rangle|_F :=& \left[ \sum_{\alpha \alpha'} |\langle \mu_\alpha^{ij}(x) \bar{\mu}^{\alpha'}_{ij}(y) \rangle|^2 \right]^{1/2}\\
\leq& C \left[\sum_{\text{All Indices}} \left| \int d\mu  \left(S_{\Psi \beta}^{\beta'}\right)_{aa'} \left(S_{\Lambda\gamma}^{\gamma'}\right)_{bc,b'c'} \left(S^{\delta'}_{\Psi \delta}\right)_{dd'}\right|^2  \right]^{1/2}
\end{split}
\nn
\ee
where $C$ is a positive constant. Importantly here we have buried the details of the colour and spinor contractions in this constant. Next we use the H\"{o}lder inequality which says for any positive definite measure
\be
\left|\int d \mu(f g)\right| \leq\left(\int d \mu|f|^{2}\right)^{\frac{1}{2}}\left(\int d \mu|g|^{2}\right)^{\frac{1}{2}}
\nn
\ee
such that
\be
\begin{split}
|\langle \mu^{ij}(x) \bar{\mu}_{ij}(y) \rangle|_F \leq& C \Bigg[ \int d\mu \sum_{\substack{\beta \beta'\\ aa'}} \left|\left(S_{\Psi \beta}^{\beta'}\right)_{aa'}\right|^2 \Bigg]^{1/2}\\ &\times \Bigg[ \int d\mu \sum_{\substack{\gamma\gamma',\delta \delta'\\bb',cc',dd'}} \left|\left(S_{\Lambda\gamma}^{\gamma'}\right)_{bc,b'c'} \left(S^{\delta'}_{\Psi \delta}\right)_{dd'}\right|^2 \Bigg]^{1/2}
\end{split}
\nn
\ee
We now want to interpret each integral, both of which are gauge invariant, as a correlation function. The first integral is equal to the propagator of the meson $\pi=\bar{\Psi}^a \gamma_5 \Psi_a$. Crucially the colour indices of $\pi$ are contracted via the the identity which allows this identification. Explicitly, in Lorentzian signature, the meson can be written in terms of Weyl fermions as $\pi=J^{ab}\psi_b \psi_a + \text{h.c.}$ as the $J$ tensors are burried in \eqref{eq:majweyl}.  The second integral can, at worst, approach a constant asymptotically. In the $|x-y| \rightarrow \infty$ limit, the mass of the lightest particle carrying the same quantum numbers as the operators dominates and we get the inequality
\be
e^{-m_\pi |x-y |}\geq e^{-m_\mu |x-y |} \;\;\;\; \Rightarrow \;\;\;\; m_\pi \leq m_\mu
\label{eq:weingartenmass}
\ee
There is no reason for the pion to be massless without symmetry breaking, so \eqref{eq:weingartenmass} prohibits $\rho$ being massless in the absence of symmetry breaking. A very similar analysis goes through for the fermionic $SO(N_c)$ theory.

This result excludes an s-confining phase where we need the massless fermion $\mu$ for anomaly matching. It also prohibits the electric to magnetic flow of the putative duals in Section \ref{spsec} if the electric theory has no scalars. This follows as the singlet $\mu$ must be massless for anomalies to match across the duality but the map $\mu = \psi \lambda \psi$ implies $\mu$ must be massive which leads to a contradiction. As explained in Section \ref{vlduals}, this is closely tied with the ``persistent mass condition".

\acknowledgments
We would like to thank Micha Berkooz, Nakarin Lohitsiri, and Tin Sulejmanpasic for discussions and correspondence on various topics. 
We are supported by the STFC consolidated grant ST/P000681/1 and the EPSRC grant EP/V047655/1 ``Chiral Gauge
Theories: From Strong Coupling to the Standard Model". KO is further supported by a Cambridge Trust International Scholarship and DT by a Simons Investigator Award. For the purpose of open access, the authors have applied a Creative Commons Attribution (CC BY) licence to any Author Accepted Manuscript version arising from this submission.

\end{document}